\pdfoutput=1
\PassOptionsToPackage{dvipsnames}{xcolor}
\documentclass[11pt,letterpaper]{article}

\newif\iffull%
\fulltrue%

\usepackage[margin=1in]{geometry}
\usepackage{hyperref}
\usepackage{subcaption}
\usepackage{array}
\usepackage{enumerate}
\usepackage{tabularx}
\usepackage{booktabs}
\usepackage{threeparttable}
\usepackage{authblk}

\usepackage{import}
\usepackage[utf8]{inputenc}

\PassOptionsToPackage{hyphens}{url}%
\usepackage{hyperref}
\usepackage{graphicx}

\usepackage{amsmath}
\usepackage{amsthm}

\usepackage{amssymb}
\usepackage{amsfonts}
\usepackage{mathtools}

\usepackage{shortsym}

\usepackage{microtype}

\usepackage{IEEEtrantools}
\allowdisplaybreaks

\usepackage[shortlabels,inline]{enumitem}
\setlist[itemize]{leftmargin=1.25em}
\setlist[enumerate]{leftmargin=1.75em}
\usepackage[normalem]{ulem}

\usepackage{import}
\theoremstyle{plain}
\newtheorem*{theorem*}{Theorem}
\newtheorem*{conjecture*}{Conjecture}

\theoremstyle{definition}
\makeatletter
\renewcommand*\env@matrix[1][*\c@MaxMatrixCols c]{%
    \hskip -\arraycolsep
    \let\@ifnextchar\new@ifnextchar
    \array{#1}}
\makeatother

\DeclarePairedDelimiter{\abs}{\lvert}{\rvert}
\DeclarePairedDelimiter{\len}{\lvert}{\rvert}
\DeclarePairedDelimiter{\norm}{\lVert}{\rVert}
\DeclarePairedDelimiter{\floor}{\lfloor}{\rfloor}
\DeclarePairedDelimiter{\ceil}{\lceil}{\rceil}

\makeatletter
\let\oldabs\abs
\def\abs{\@ifstar{\oldabs}{\oldabs*}}
\let\oldlen\len
\def\len{\@ifstar{\oldlen}{\oldlen*}}
\let\oldnorm\norm
\def\norm{\@ifstar{\oldnorm}{\oldnorm*}}
\let\oldfloor\floor
\def\floor{\@ifstar{\oldfloor}{\oldfloor*}}
\let\oldceil\ceil
\def\ceil{\@ifstar{\oldceil}{\oldceil*}}
\makeatother

\usepackage{pifont}
\newcommand{\cmark}{\ding{52}}%
\newcommand{\xmark}{\ding{56}}%

\usepackage{dsfont}

\usepackage{xspace}

\usepackage[dvipsnames]{xcolor}

\definecolor{myA16zGrayLight}{RGB}{235,235,235}     %
\definecolor{myA16zGrayMedium}{RGB}{196,196,196}    %
\definecolor{myA16zGrayDark}{RGB}{44,34,34}         %
\definecolor{myA16zLavender}{RGB}{208,161,255}      %
\definecolor{myA16zMagenta}{RGB}{195,70,206}        %
\definecolor{myA16zMulberry}{RGB}{113,24,88}        %
\definecolor{myA16zLemonChiffon}{RGB}{250,234,157}  %
\definecolor{myA16zAmber}{RGB}{230,154,48}          %
\definecolor{myA16zRust}{RGB}{174,59,10}            %
\definecolor{myA16zLime}{RGB}{197,222,107}          %
\definecolor{myA16zAquamarine}{RGB}{82,216,145}     %
\definecolor{myA16zPine}{RGB}{60,87,44}             %
\definecolor{myA16zPacific}{RGB}{145,224,235}       %
\definecolor{myA16zTeal}{RGB}{36,197,201}           %
\definecolor{myA16zAzure}{RGB}{18,51,90}            %

\definecolor{myTechnionDeepBlue}{HTML}{002147}           %
\definecolor{myTechnionGoldenOchre}{HTML}{D59F0F}        %
\definecolor{myTechnionBlack}{HTML}{000000}              %
\definecolor{myTechnionWhite}{HTML}{FFFFFF}              %

\definecolor{myTechnionRed}{HTML}{E31D1A}             %
\definecolor{myTechnionPink}{HTML}{EA094B}           %
\definecolor{myTechnionPurple1}{HTML}{AE3B72}         %
\definecolor{myTechnionPurple2}{HTML}{4D4084}         %
\definecolor{myTechnionBlue1}{HTML}{216093}            %
\definecolor{myTechnionBlue2}{HTML}{5686DA}           %
\definecolor{myTechnionTeal}{HTML}{32B1CA}            %
\definecolor{myTechnionGreen1}{HTML}{EA094B}           %
\definecolor{myTechnionGreen2}{HTML}{A3D65C}           %
\definecolor{myTechnionGreen3}{HTML}{94D60A}           %
\definecolor{myTechnionYellow}{HTML}{FDD700}          %
\definecolor{myTechnionOrange}{HTML}{FF6B00}         %
\definecolor{myTechnionBrown}{HTML}{97775C}          %
\definecolor{myTechnionBeige}{HTML}{D9D1C3}          %
\definecolor{myTechnionGray1}{HTML}{A2A9AE}            %
\definecolor{myTechnionGray2}{HTML}{5A6771}            %

\definecolor{mySuCardinalRed}{HTML}{8c1515}
\definecolor{mySuCardinalRedLight}{HTML}{B83A4B}
\definecolor{mySuCardinalRedDark}{HTML}{820000}
\definecolor{mySuWhite}{HTML}{ffffff}
\definecolor{mySuCoolGrey}{HTML}{53565A}
\definecolor{mySuBlack}{HTML}{2e2d29}
\definecolor{mySuBlack100}{HTML}{2e2d29}
\definecolor{mySuBlack90}{HTML}{43423E}
\definecolor{mySuBlack80}{HTML}{585754}
\definecolor{mySuBlack70}{HTML}{6D6C69}
\definecolor{mySuBlack60}{HTML}{767674}
\definecolor{mySuBlack50}{HTML}{979694}
\definecolor{mySuBlack40}{HTML}{ABABA9}
\definecolor{mySuBlack30}{HTML}{C0C0BF}
\definecolor{mySuBlack20}{HTML}{D5D5D4}
\definecolor{mySuBlack10}{HTML}{EAEAEA}

\definecolor{mySuPaloAlto}{HTML}{175E54}
\definecolor{mySuPaloAltoLight}{HTML}{2D716F}
\definecolor{mySuPaloAltoDark}{HTML}{014240}
\definecolor{mySuPaloVerde}{HTML}{279989}
\definecolor{mySuPaloVerdeLight}{HTML}{59B3A9}
\definecolor{mySuPaloVerdeDark}{HTML}{017E7C}
\definecolor{mySuOlive}{HTML}{8F993E}
\definecolor{mySuOliveLight}{HTML}{A6B168}
\definecolor{mySuOliveDark}{HTML}{7A863B}
\definecolor{mySuBay}{HTML}{6FA287}
\definecolor{mySuBayLight}{HTML}{8AB8A7}
\definecolor{mySuBayDark}{HTML}{417865}
\definecolor{mySuSky}{HTML}{4298B5}
\definecolor{mySuSkyLight}{HTML}{67AFD2}
\definecolor{mySuSkyDark}{HTML}{016895}
\definecolor{mySuLagunita}{HTML}{007C92}
\definecolor{mySuLagunitaLight}{HTML}{009AB4}
\definecolor{mySuLagunitaDark}{HTML}{006B81}
\definecolor{mySuPoppy}{HTML}{E98300}
\definecolor{mySuPoppyLight}{HTML}{F9A44A}
\definecolor{mySuPoppyDark}{HTML}{D1660F}
\definecolor{mySuSpirited}{HTML}{E04F39}
\definecolor{mySuSpiritedLight}{HTML}{F4795B}
\definecolor{mySuSpiritedDark}{HTML}{C74632}
\definecolor{mySuIlluminating}{HTML}{FEDD5C}
\definecolor{mySuIlluminatingLight}{HTML}{FFE781}
\definecolor{mySuIlluminatingDark}{HTML}{FEC51D}
\definecolor{mySuPlum}{HTML}{620059}
\definecolor{mySuPlumLight}{HTML}{734675}
\definecolor{mySuPlumDark}{HTML}{350D36}
\definecolor{mySuBrick}{HTML}{651C32}
\definecolor{mySuBrickLight}{HTML}{7F2D48}
\definecolor{mySuBrickDark}{HTML}{42081B}
\definecolor{mySuArchway}{HTML}{5D4B3C}
\definecolor{mySuArchwayLight}{HTML}{766253}
\definecolor{mySuArchwayDark}{HTML}{2F2424}
\definecolor{mySuStone}{HTML}{7F7776}
\definecolor{mySuStoneLight}{HTML}{D4D1D1}
\definecolor{mySuStoneDark}{HTML}{544948}
\definecolor{mySuFog}{HTML}{DAD7CB}
\definecolor{mySuFogLight}{HTML}{F4F4F4}
\definecolor{mySuFogDark}{HTML}{B6B1A9}

\definecolor{mySuDigitalRed}{HTML}{B1040E}
\definecolor{mySuDigitalRedLight}{HTML}{E50808}
\definecolor{mySuDigitalRedDark}{HTML}{820000}
\definecolor{mySuDigitalBlue}{HTML}{006CB8}
\definecolor{mySuDigitalBlueLight}{HTML}{6FC3FF}
\definecolor{mySuDigitalBlueDark}{HTML}{00548f}
\definecolor{mySuDigitalGreen}{HTML}{008566}
\definecolor{mySuDigitalGreenLight}{HTML}{1AECBA}
\definecolor{mySuDigitalGreenDark}{HTML}{006F54}

\definecolor{myParula1Blue}{RGB}{0,114,189}
\definecolor{myParula2Orange}{RGB}{217,83,25}
\definecolor{myParula3Yellow}{RGB}{237,177,32}
\definecolor{myParula4Purple}{RGB}{126,47,142}
\definecolor{myParula5Green}{RGB}{119,172,48}
\definecolor{myParula6LightBlue}{RGB}{77,190,238}
\definecolor{myParula7Red}{RGB}{162,20,47}

\usepackage{multirow}
\usepackage{booktabs}   %
\usepackage{makecell}
\usepackage{threeparttable}   %

\usepackage{algorithm}
\usepackage{algorithmicx}
\usepackage[noend]{algpseudocode}

\makeatletter
\AddToHook{env/algorithmic/begin}{\def\@currentcounter{ALG@line}}
\renewcommand{\theHALG@line}{\thealgorithm.\arabic{ALG@line}}
\makeatother

\algnewcommand{\LineComment}[1]{\State {\textcolor{gray}{/\!/ #1}}}

\algrenewcommand{\alglinenumber}[1]{\scriptsize\textcolor{gray}{\texttt{#1}}}
\algrenewcommand{\algorithmicindent}{1em}

\algnewcommand{\algfontsize}[0]{}
\AddToHook{env/algorithmic/begin}{\algfontsize}

\algnewcommand{\algorithmicswitch}{\textbf{switch}}
\algdef{SE}[SWITCH]{Switch}{EndSwitch}[1]{\algorithmicswitch\ #1\ \algorithmicdo}{\algorithmicend\ \algorithmicswitch}%
\algtext*{EndSwitch}%

\algnewcommand{\algorithmiccase}{\textbf{case}}
\algdef{SE}[CASE]{Case}{EndCase}[1]{\algorithmiccase\ #1}{\algorithmicend\ \algorithmiccase}%
\algtext*{EndCase}%

\algnewcommand{\algorithmicon}{\textbf{on}}
\algdef{SE}[ON]{On}{EndOn}[1]{\algorithmicon\ #1\ \algorithmicdo}{\algorithmicend\ \algorithmicon}%
\algtext*{EndOn}%

\algnewcommand{\algorithmicupon}{\textbf{upon}}
\algdef{SE}[UPON]{Upon}{EndUpon}[1]{\algorithmicupon\ #1\ \algorithmicdo}{\algorithmicend\ \algorithmicupon}%
\algtext*{EndUpon}%

\algnewcommand{\algorithmicduring}{\textbf{during}}
\algdef{SE}[DURING]{During}{EndDuring}[1]{\algorithmicduring\ #1\ \algorithmicdo}{\algorithmicend\ \algorithmicduring}%
\algtext*{EndDuring}%

\algnewcommand{\algorithmicat}{\textbf{at}}
\algdef{SE}[AT]{At}{EndAt}[1]{\algorithmicat\ #1\ \algorithmicdo}{\algorithmicend\ \algorithmicat}%
\algtext*{EndAt}%

\algnewcommand{\algorithmicrealfunction}{\textbf{function}}
\algdef{SE}[REALFUNCTION]{RealFunction}{EndRealFunction}[1]{\algorithmicrealfunction\ #1\ \algorithmicdo}{\algorithmicend\ \algorithmicrealfunction}%
\algtext*{EndRealFunction}%

\algnewcommand{\algorithmicthroughout}{\textbf{do throughout}}
\algdef{SE}[Throughout]{Throughout}{EndThroughout}[1]{\algorithmicthroughout\ #1\ \algorithmicdo}{\algorithmicend\ \algorithmicthroughout}%
\algtext*{EndThroughout}%

\algnewcommand{\algorithmictry}{\textbf{try}}
\algnewcommand{\algorithmiccatch}{\textbf{catch}}
\algdef{SE}[TRY]{Try}{EndTry}{\algorithmictry\ \algorithmicdo}{\algorithmicend\ \algorithmictry}
\algdef{C}[TRY]{TRY}{Catch}[1]{\algorithmiccatch\ #1\ \algorithmicdo}
\algtext*{EndTry}

\algrenewcommand{\algorithmicdo}{}
\algrenewcommand{\algorithmicthen}{}

\algnewcommand{\algorithmicgoto}{\textbf{goto}}%
\algnewcommand{\Goto}[1]{\algorithmicgoto~\ref{#1}}%

\algnewcommand{\algorithmicassert}{\textbf{assert}}%
\algnewcommand{\Assert}[1]{\algorithmicassert~{#1}}%

\algnewcommand{\algorithmicbreak}{\textbf{break}}%
\algnewcommand{\Break}[0]{\algorithmicbreak}%
\algnewcommand{\BreakOutOf}[1]{\algorithmicbreak~out~of~#1}%

\algnewcommand{\algorithmicwaiton}{\textbf{wait on}}%
\algnewcommand{\WaitOn}[1]{\algorithmicwaiton~{#1}}%

\algnewcommand{\InlineRequire}[1]{\textbf{require} {#1}}

\algblock{ManualIndent}{EndManualIndent}
\algnotext{ManualIndent}
\algnotext{EndManualIndent}

\algdef{SE}[GENERICBLOCK]{GenericBlock}{EndGenericBlock}[1]{#1}{}%
\algtext*{EndGenericBlock}%

\usepackage[sort&compress,capitalize,nameinlink]{cleveref}

\AtBeginEnvironment{appendices}{%
    \crefalias{section}{appendix}%
    \crefalias{subsection}{subappendix}%
    \crefalias{subsubsection}{subsubappendix}%
    \crefalias{subsubsubsection}{subsubsubappendix}%
}
\AddToHook{cmd/appendix/after}{%
    \crefalias{section}{appendix}%
    \crefalias{subsection}{subappendix}%
    \crefalias{subsubsection}{subsubappendix}%
    \crefalias{subsubsubsection}{subsubsubappendix}%
}

\hypersetup{hypertexnames=false}

\crefalias{ALG@line}{line}
\usepackage{tikz}
\usetikzlibrary{fit}
\usetikzlibrary{math}
\usetikzlibrary{calc}
\usetikzlibrary{positioning}
\usetikzlibrary{decorations.pathmorphing}
\usetikzlibrary{decorations.pathreplacing}
\usetikzlibrary{calligraphy}
\usetikzlibrary{backgrounds}
\usetikzlibrary{patterns}
\usetikzlibrary{matrix}

\newcommand{\ccsnormalsize}{\fontsize{9}{11}\selectfont}

\makeatletter
\NewDocumentCommand {\getnodedimen} {O{\nodewidth} O{\nodeheight} m} {
    \begin{pgfinterruptboundingbox}
        \begin{scope}[local bounding box=bb@temp]
            \node[inner sep=0pt, fit=(#3)] {};
        \end{scope}
        \path ($(bb@temp.north east)-(bb@temp.south west)$);
    \end{pgfinterruptboundingbox}
    \pgfgetlastxy{#1}{#2}
}
\makeatother

\AtBeginDocument{\sbox0{$x$}}%
\tikzset{baseshift/.style={yshift=-\the\dimexpr\fontdimen22\textfont2}}
\tikzset{
    MAT/.style={ampersand replacement=\&},
    XS/.style={minimum size=3mm, font=\ccsnormalsize, inner sep=0pt, text width=3mm, align=center},
    SM/.style={minimum size=4.5mm, font=\ccsnormalsize, inner sep=0pt, text width=4.5mm, align=center},
    MD/.style={scale=1.25},
    LG/.style={scale=1.5},
    XL/.style={scale=2},
    MEMBER/.style={draw, circle, very thick, minimum size=6mm, inner sep=0pt, text width=6mm, font=\ccsnormalsize, align=center},
    ADV/.style={draw=red!60, text=red},
    ASLEEP/.style={draw=gray!60, text=gray, densely dotted},
    BOOTING/.style={draw=black!70, dashed},
    NONMEMBER/.style={draw=red!70!gray, text=red!70!gray, dashed},
    MEMSET/.style={draw, rounded corners=5pt, inner xsep=1mm, inner ysep=1mm},
    ESTMEMSET/.style={MEMSET, dashed, rounded corners=5pt, inner xsep=1mm, inner ysep=0.25mm},
    BLACKBOX/.style={draw=black, rectangle, thick, minimum width=8mm, minimum height=6mm,
            font=\large\color{white}, fill=black, rounded corners=2pt},
    RECON/.style={->, line width=0.5mm},
    ADVRECON/.style={->, line width=0.5mm, red, dashed},
    TX/.style={->, line width=0.5mm},
    ADVTX/.style={->, line width=0.5mm, red, dotted},
    REQ/.style={-latex, draw=black!70},
    GOODBOX/.style={draw=none, fill=green!80!black!50, opacity=0.2, rounded corners=5pt, inner xsep=4mm, inner ysep=2mm},
    BADBOX/.style={draw=none, fill=red!75, opacity=0.2, rounded corners=5pt, inner xsep=4mm, inner ysep=2mm},
    RANGE/.style={-latex, draw=red!60, opacity=0.75, line width=0.25mm},
    SIM/.style={-, draw=red!60, opacity=0.75, line width=0.25mm, decorate, decoration={snake, amplitude=0.25mm, segment length=3mm}},
    CALLOUT/.style={draw, thick, rounded corners, font=\ccsnormalsize, align=center},
    PURPLEMSG/.style={fill=purple!20},
    ORANGEMSG/.style={fill=orange!20},
    BLUEMSG/.style={fill=cyan!20},
    GREENMSG/.style={fill=green!80!black!20},
    SIMMSG/.style={draw=red!60, dotted},
}
\subimport{./lib/}{defer.tex}

\providecommand{\Description}[1]{}
\theoremstyle{plain}
\newtheorem{theorem}{Theorem}[section]
\theoremstyle{definition}
\newtheorem{definition}[theorem]{Definition}

\usepackage[sort&compress,capitalize,nameinlink]{cleveref}

\crefalias{ALG@line}{line}

\crefname{figure}{Fig.}{Figs.}
\Crefname{figure}{Fig.}{Figs.}

\crefname{table}{Tab.}{Tabs.}
\Crefname{table}{Tab.}{Tabs.}

\crefname{section}{Sec.}{Secs.}
\Crefname{section}{Sec.}{Secs.}
\crefname{subsection}{Sec.}{Secs.}
\Crefname{subsection}{Sec.}{Secs.}
\crefname{subsubsection}{Sec.}{Secs.}
\Crefname{subsubsection}{Sec.}{Secs.}
\crefname{subsubsubsection}{Sec.}{Secs.}
\Crefname{subsubsubsection}{Sec.}{Secs.}
\crefname{appendix}{App.}{Apps.}
\Crefname{appendix}{App.}{Apps.}
\crefname{subappendix}{App.}{Apps.}
\Crefname{subappendix}{App.}{Apps.}
\crefname{subsubappendix}{App.}{Apps.}
\Crefname{subsubappendix}{App.}{Apps.}
\crefname{subsubsubappendix}{App.}{Apps.}
\Crefname{subsubsubappendix}{App.}{Apps.}

\crefname{algorithm}{Alg.}{Algs.}
\Crefname{algorithm}{Alg.}{Algs.}
\crefname{line}{ln.}{lns.}
\Crefname{line}{ln.}{lns.}

\crefname{proposition}{Prop.}{Props.}
\Crefname{proposition}{Prop.}{Props.}
\crefname{lemma}{Lem.}{Lems.}
\Crefname{lemma}{Lem.}{Lems.}
\crefname{theorem}{Thm.}{Thms.}
\Crefname{theorem}{Thm.}{Thms.}
\crefname{corollary}{Cor.}{Cors.}
\Crefname{corollary}{Cor.}{Cors.}
\crefname{definition}{Def.}{Defs.}
\Crefname{definition}{Def.}{Defs.}
\crefname{conjecture}{Conj.}{Conjs.}
\Crefname{conjecture}{Conj.}{Conjs.}
\crefname{remark}{Rem.}{Rems.}
\Crefname{remark}{Rem.}{Rems.}

\usepackage{xspace}

\MakeRobust{\Call}

\newcommand{\code}[1]{\Call{#1}{}}

\newcolumntype{Y}{>{\centering\arraybackslash}X}
\newcolumntype{Z}{>{\raggedright\arraybackslash}X}

\newcommand{\bad}{\widehat}
\DeclareRobustCommand{\honpic}{\tikz[baseline=(current bounding box.base)]{\node[MEMBER, XS, yshift=3] {};}\xspace}
\DeclareRobustCommand{\dapic}{\tikz[baseline=(current bounding box.base)]{\node[MEMBER, XS, ASLEEP, yshift=3] {};}\xspace}
\DeclareRobustCommand{\advpic}{\tikz[baseline=(current bounding box.base)]{\node[MEMBER, XS, ADV, yshift=3] {};}\xspace}
\DeclareRobustCommand{\hadvpic}{\tikz[baseline=(current bounding box.base)]{\node[MEMBER, NONMEMBER, XS, yshift=3] {};}\xspace}
\DeclareRobustCommand{\sleepadvpic}{\tikz[baseline=(current bounding box.base)]{\node[MEMBER, XS, ASLEEP, ADV, yshift=3] {};}\xspace}
\DeclareRobustCommand{\longrangepic}{\tikz[baseline=(current bounding box.base)]{\draw[RANGE,yshift=3] (0.75,0) to[out=160,in=15] (0,0) {};}\xspace}
\DeclareRobustCommand{\simpic}{\tikz[baseline=(current bounding box.base)]{\draw[SIM] (0,0) -- (0.5,0) {};\draw[SIM] (0,0.2) -- (0.5,0.2) {};}\xspace}

\DeclareRobustCommand{\txpic}{\tikz[baseline=(current bounding box.base)]{\draw[TX,yshift=3] (0,0) -- (0.5,0) {};}\xspace}
\DeclareRobustCommand{\advtxpic}{\tikz[baseline=(current bounding box.base)]{\draw[ADVTX,yshift=3] (0,0) -- (0.5,0) {};}\xspace}
\DeclareRobustCommand{\reqpic}{\tikz[baseline=(current bounding box.base)]{\draw[REQ,yshift=3] (0,0) -- (0.5,0) {};}\xspace}
\DeclareRobustCommand{\advreconpic}{\tikz[baseline=(current bounding box.base)]{\draw[ADVRECON,yshift=3] (0,0) -- (0.5,0) {};}\xspace}

\newcommand{\protocol}{\Pi}
\newcommand{\da}{\protocol^{\mathsf{DA}}}
\newcommand{\eda}[1][\epoch]{\da_{#1}}

\newcommand{\hon}{\mathcal{H}}
\newcommand{\hons}[3][]{\mathcal{H}^{#1}(#2, #3)}
\newcommand{\adv}{\mathcal{A}}
\newcommand{\sims}[3][]{\mathcal{S}^{#1}(#2; #3)}
\newcommand{\equivs}{\mathcal{D}}
\newcommand{\withdrawer}{\mathcal{W}}
\newcommand{\withdrawers}[3][]{\withdrawer^{#1}(#2, #3)}

\newcommand{\udsims}[3][]{\mathcal{S}_{\overline{\withdrawer}}^{#1}(#2; #3)}

\newcommand{\exec}{\mathsf{X}}
\newcommand{\schd}{\mathsf{K}}
\newcommand{\cond}{\mathsf{C}}
\newcommand{\nodemap}{\phi}

\newcommand{\chain}{\Lambda}
\newcommand{\echain}[1][\epoch]{\chain_{#1}}

\newcommand{\mem}{\mathcal{M}}
\newcommand{\memest}{\mathsf{E}}

\newcommand{\genesis}{\mem_0}

\newcommand{\txs}{\mathcal{T}}

\newcommand{\conflicts}{\mathcal{C}}

\newcommand{\hm}{HM\xspace}

\newcommand{\sr}{SR-HM\xspace}

\newcommand{\srhfull}{\sr condition (\cref{def:srhm})\xspace}
\newcommand{\srfull}{the \srhfull}

\newcommand{\srhfulllight}{\sr condition\xspace}
\newcommand{\srfulllight}{the \srhfulllight}

\newcommand{\nrounds}{R}
\newcommand{\epoch}{e}
\newcommand{\rnd}{t}
\newcommand{\srnd}{s}

\newcommand{\emem}[1][\epoch]{\tilde\mem(#1)}
\newcommand{\getmem}[1]{#1.\Call{newMembers}{ }}
\newcommand{\gettxs}[1]{#1.\mathsf{txs}}
\newcommand{\applytxs}[2]{\Call{ApplyTxs}{#1, #2}}

\newcommand{\msgs}{\mathsf{msgs}}
\newcommand{\votemsgtype}{\mathsf{MVOTE}}
\newcommand{\votemsg}[1]{(\votemsgtype, #1)}
\newcommand{\logmsgtype}{\mathsf{LVOTE}}
\newcommand{\logmsg}[1]{(\logmsgtype, #1)}
\newcommand{\txmsgtype}{\mathsf{TX}}
\newcommand{\txmsg}[1]{(\txmsgtype, #1)}

\newcommand{\votes}{\mathsf{votes}}

\newcommand{\memvar}{\mathsf{M}}

\newcommand{\crypto}{\mathcal{F}}
\newcommand{\kes}{\crypto_\code{FS}}

\newcommand{\sk}{\code{sk}}
\newcommand{\pk}{\code{pk}}

\newcommand{\signed}[2][]{\left\langle #2 \right\rangle^{#1}}
\newcommand{\signedt}[3]{\signed[#2]{#1}_{#3}}
\newdefergroup{proof}[notdeferred]

\title{On the Limits of Consensus\texorpdfstring{\\}{ }under Dynamic Availability and Reconfiguration}

\author[1]{Javier Nieto}
\author[2]{Joachim Neu}
\author[1]{Ling Ren}
\affil[1]{University of Illinois Urbana-Champaign, \texttt{\{jmnieto2,renling\}@illinois.edu}}
\affil[2]{a16z Crypto Research, \texttt{jneu@a16z.com}}
\date{}

\crefname{theorem}{Thm.}{Thms.}
\Crefname{theorem}{Thm.}{Thms.}
\crefname{definition}{Def.}{Defs.}
\Crefname{definition}{Def.}{Defs.}

\begin{document}%
\maketitle%
\begin{abstract}
    Proof-of-stake blockchains require consensus protocols that support Dynamic Availability and Reconfiguration (which we call the DAR model).
    Here, Dynamic Availability means that the consensus protocol should remain live even if a large number of nodes temporarily crash, and Reconfiguration means it should be possible to change the set of operating nodes over time.
    State-of-the-art protocols inspired by the DAR model, such as Ethereum, Cardano's Ouroboros, or Snow White, require additional
    model features
    comprising
    external mechanisms or node capabilities
    that
    are difficult to justify, such as social consensus or the requirement that key evolution be performed even when nodes have crashed.
    The key result of this paper is the necessary and sufficient adversarial condition under which consensus can be achieved in the \emph{plain} DAR model, \emph{without any extra features}.
    We then introduce an additional feature to the model that is justified by proof-of-stake blockchain designs:
    honest nodes complete a sign-off procedure the moment they express intent to exit from the set of operating nodes.
    This additional feature reduces the power of the adversary relative to the plain DAR model and helps us obtain a bootstrapping gadget that is particularly simple and efficient in the common optimistic case of few reconfigurations and no double spending.
\end{abstract}

\section{Introduction}
\label{intro}

Today's distributed systems are evolving from static, tightly coordinated deployments into open, decentralized environments with dynamic availability and reconfigurable membership.
For instance, classical consensus protocols such as PBFT~\cite{CL99} assume a fixed set of always-online nodes (also called servers, replicas, or validators).
This is in stark contrast to modern proof-of-stake (PoS) blockchains, where nodes may frequently go offline, join, or leave the system.
To this end, current literature has studied two orthogonal dimensions of changing participation.

First, the idea of \emph{reconfigurable} membership in distributed systems came early on to support adding, removing, or replacing nodes after initialization of the protocol.
Whereas classical consensus is defined on a single fixed \emph{membership set}, reconfigurable systems consist of a sequence of membership sets where a set of nodes makes consensus decisions for some time before the next set of nodes takes over~\cite{LMZ09,LMZ10,SRMJ12,OO14,BSA14,JM14}.

Second, and orthogonal to reconfiguration, is the idea of \emph{dynamic availability}, which was pioneered by the Bitcoin protocol~\cite{Nak09}. %
Bitcoin does not require a fixed quorum of nodes to make decisions.
Its fault tolerance is thus inherently determined by the relative levels of participation, i.e., the proportion of honest versus adversarial computational power, at any given time.
As a result, Bitcoin remains operational even if a large number of nodes temporarily stop participating. %
Pass and Shi~\cite{PS17} first formalized this notion as the \emph{sleepy model}, which is also referred to as the dynamic availability model in the literature~\cite{NTT20a,DNTT23}.
The model considers a \emph{fixed} set of nodes, where, throughout execution, some honest nodes may become \emph{asleep}
and not participate in the protocol.
Asleep nodes can be viewed as crashed nodes that may recover later.
However, the sleepy model is not to be confused with mixed fault models~\cite{GP92,CMSK07,ADKS22,GDCL22}, because the sleepy model allows for an arbitrary number of asleep nodes, whereas mixed fault models put a bound on the total number of Byzantine and crash-recovery faults.

\begin{table*}[t!]
    \caption{A comparison of protocols in the DAR model and its variants}
    \label{tbl:assumptions}
    \centering
    \iffull\small\fi
    \def\tblassumptionsrows{%
        \toprule
        \textbf{Protocol}               & \textbf{Model}                    & \textbf{Adversarial condition}                                & \textbf{No external mechanism} & \textbf{No added node capability} \\
        \midrule
        Bitcoin~\cite{Nak09}            & PoW                               & Honest majority                                               & \textcolor{Green}{\cmark}      & \textcolor{Green}{\cmark}         \\
        \midrule
        Snow White~\cite{DPS16}         & DAR with social consensus         & Honest majority                                               & \textcolor{Red}{\xmark}        & \textcolor{Green}{\cmark}         \\
        Ethereum~\cite{BHK+20}          & DAR with social consensus         & Honest majority                                               & \textcolor{Red}{\xmark}        & \textcolor{Green}{\cmark}         \\
        Ouroboros Genesis~\cite{BGK+18} & DAR with continuous key evolution & Honest majority                                               & \textcolor{Green}{\cmark}      & \textcolor{Red}{\xmark}           \\
        \midrule
        \cref{alg:dar-gadget}           & DAR                               & SR-HM (\cref{def:srhm,def:simulatable})\tnote{*}              & \textcolor{Green}{\cmark}      & \textcolor{Green}{\cmark}         \\
        \cref{alg:darso-gadget}         & DAR with sign-off                 & SR-HM (\cref{def:srhm,def:non-disposed-simulatable})\tnote{*} & \textcolor{Green}{\cmark}      & \textcolor{Red}{\xmark}           \\
        \bottomrule
    }%
    \begin{threeparttable}
        \iffull
            \begin{tabularx}{\textwidth}{>{\hsize=1.05\hsize}Z>{\hsize=1.15\hsize}Z>{\hsize=1.50\hsize}Z>{\hsize=0.7\hsize}Y>{\hsize=0.6\hsize}Y}
                \tblassumptionsrows
            \end{tabularx}
        \else
            \begin{tabularx}{\textwidth}{l>{\hsize=0.4\hsize}l>{\hsize=0.4\hsize}l>{\hsize=0.4\hsize}Y>{\hsize=0.4\hsize}Y}
                \tblassumptionsrows
            \end{tabularx}
        \fi
        \begin{tablenotes}
            \item[*] SR-HM stands for simulation-resistant honest majority. The definition of simulation-resistant differs slightly under the plain DAR model and the DAR with sign-off model (see \cref{def:srhm,def:non-disposed-simulatable,def:simulatable}).
        \end{tablenotes}
    \end{threeparttable}
\end{table*}

It is then natural to combine the two orthogonal aspects into the \emph{Dynamic Availability and Reconfiguration} (DAR) model.
While prior works have explored models that appear similar to DAR~\cite{DPS16,BGK+18,BHK+20}, a closer inspection reveals that these works incorporate additional external mechanisms or node capabilities into the model that are difficult to justify, as we elaborate below.
Consequently, these works do not operate in what we call, for clarity, the \emph{plain} DAR model, but rather in \emph{variants} of the DAR model \emph{with extra features}.

Snow White~\cite{DPS16} and PoS Ethereum~\cite{BHK+20} rely on an extra external mechanism called \emph{social consensus}, which states that newly joining nodes obtain a recent checkpoint of the system through some mechanism.
In practice, this mechanism must continuously achieve consensus on an up-to-date sequence of checkpoints so that outdated nodes can determine the latest state to correctly participate in the consensus protocol.
The common rationale for social consensus is that newly joining nodes must anyway obtain a correct protocol implementation from some trusted source; thus, obtaining a recent checkpoint of the blockchain state appears to impose little additional requirement.
However, this seemingly minor requirement is at odds with the core principles and established motivation of distributed consensus.
While all consensus protocols require some initial common knowledge~\cite{HM00} (i.e., agreed-upon states), such as the protocol description itself and a few public parameters (like the genesis block in Bitcoin), the role of the consensus protocol is to subsequently process impromptu inputs and to continuously expand the agreed-upon state.
Social consensus is an external mechanism that provides consensus on a growing sequence of checkpoints to external observers.
It therefore seems unsatisfactory, or even circular, for a consensus protocol to assume the existence of an external consensus mechanism in order to solve consensus.

Ouroboros Genesis~\cite{BGK+18} is a longest-chain consensus protocol that appears to be designed for the DAR model\footnote{Ouroboros Genesis~\cite{BGK+18} uses the term dynamic availability to refer to what we define as the DAR model. We use the term dynamic availability to only refer to the availability/sleepy status of nodes, and we treat reconfiguration (membership changes) as a separate aspect of the model.}, and does not rely on social consensus.
However, their model deviates from the plain DAR model by requiring asleep nodes to \emph{continuously} perform \emph{key evolution} to maintain forward security.
Asleep nodes that fail to do so are counted as adversarial.
However, this extra capability of asleep nodes seems to be at odds with the motivation of dynamic availability.
The main goal of dynamic availability is to model nodes (stakeholders) that may not be interested in taking part in running the consensus protocol.
It seems questionable to assume that these nodes \emph{continuously} update their local state despite not actively participating in the protocol.

Thus, existing protocols achieve consensus only in variants of the DAR model that are augmented with powerful extra features that are hard to justify. %
This leaves open the question of under what condition consensus can be achieved in the \emph{plain} DAR model.\footnote{%
      One could run the protocols from prior works~\cite{DPS16,BGK+18,BHK+20} in the plain DAR model rather than their intended models and investigate the conditions under which they achieve consensus.
      We do not pursue this direction, as it is unlikely to uncover the tight condition.}
In this work, as the first key contribution, we resolve the open question, and precisely characterize the \emph{adversarial condition} that is both necessary and sufficient for achieving consensus in the plain DAR model.
Carrying the classic threshold adversary to the DAR model, the adversarial condition covers the fraction of adversarial nodes, as well as nodes' sleep schedule and the reconfiguration schedule.
In the process, we rigorously characterize and tackle the core difficulty of the DAR model: the uncertainty in who \emph{can} participate (reconfiguration) and who \emph{does} participate (availability).
Prior works dodge this difficulty by relying on extra model features.
Concretely, social consensus alleviates the uncertainty in who can participate by providing outdated nodes with oracle access to a recent state of the system, and key evolution while asleep alleviates the uncertainty in who has participated by ensuring that compromised keys cannot be used to retroactively forge participation for asleep nodes.

In the second half of this work, we propose our own feature to add to the plain DAR model:
nodes are required, one time only, to execute a \emph{sign-off} procedure when exiting the membership set,
even if the node is asleep.
While this can be regarded as an extra node capability, we contend it is realistic and representative of PoS blockchains where nodes leave the membership set by signing a transaction to transfer their stake to another node.
We refer to this stronger variant of the DAR model as the \emph{DAR with sign-off} model.
Under this model, the power of the adversary is weakened, which allows consensus to be achieved under a weaker adversarial condition than the one required in the plain DAR model.
We also give a protocol with much better efficiency than the one in the plain DAR model when only a small fraction of members is reconfigured per epoch and no double-spending transfers of stake are observed, which is often the case in real-world PoS blockchains.

We compare our models, adversarial condition, and protocols with those of the aforementioned prior works in \cref{tbl:assumptions}.

\paragraph*{Main Results.}
\begin{itemize}

      \item We formalize the necessary adversarial condition for solving consensus in the plain DAR model as \emph{simulation-resistant honest majority} (\sr, \cref{def:srhm}, see \cref{thm:srhm-necessary}).

      \item We propose a generic bootstrapping gadget that transforms any consensus protocol secure in the dynamically available and non-reconfigurable model into a consensus protocol secure in the plain DAR model under \srfulllight (see \cref{thm:dar-gadget}), thereby establishing it as the sufficient adversarial condition for solving consensus in the plain DAR model.

      \item We introduce the \emph{DAR with sign-off} model and propose a generic bootstrapping gadget with a simple and efficient good-case path that leverages signed membership-change transactions to achieve consensus in the DAR with sign-off model (see \cref{thm:darso-gadget}).
\end{itemize}

\subsection{Overview}
\label{intro-overview}

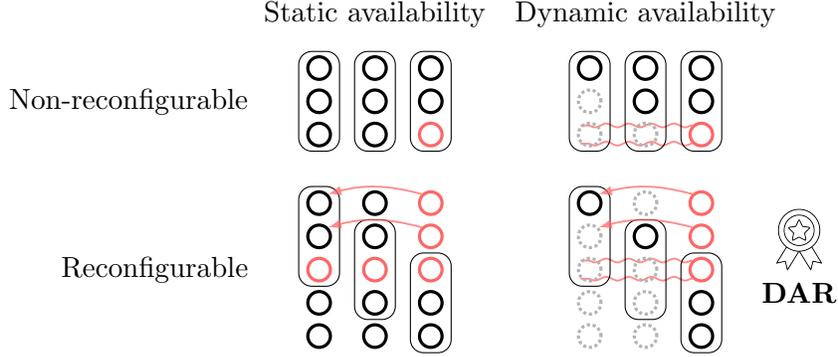
\begin{figure*}[tbp!]
      \centering
      \begin{tikzpicture}[node distance=4mm]
            \def\matColSep{4}
            \def\matRowSep{2}

            \matrix[MAT,
                  matrix of nodes,
                  column sep=\matColSep mm,
                  row sep=\matRowSep mm,
                  nodes={MEMBER, XS, anchor=center},
            ] (static) {
                  \node (p0_0) {}; \&
                  \node (p0_1) {}; \&
                  \node (p0_2) {};      \\
                  \node (p1_0) {}; \&
                  \node (p1_1) {}; \&
                  \node (p1_2) {};      \\
                  \node (p2_0) {}; \&
                  \node (p2_1) {}; \&
                  \node[ADV] (p2_2) {}; \\
            };

            \node [fit=(p0_0) (p2_0), MEMSET] {};

            \node [fit=(p0_1) (p2_1), MEMSET] {};

            \node [fit=(p0_2) (p2_2), MEMSET] {};

            \matrix[MAT,
                  matrix of nodes,
                  column sep=\matColSep mm,
                  row sep=\matRowSep mm,
                  right=1.5cm of static,
                  nodes={MEMBER, XS, anchor=center},
            ] (da) {
                  \node (p0_0) {}; \&
                  \node (p0_1) {}; \&
                  \node (p0_2) {};      \\
                  \node[ASLEEP] (p1_0) {}; \&
                  \node (p1_1) {}; \&
                  \node (p1_2) {};      \\
                  \node[ASLEEP] (p2_0) {}; \&
                  \node[ASLEEP] (p2_1) {}; \&
                  \node[ADV] (p2_2) {}; \\
            };

            \node [fit=(p0_0) (p2_0), MEMSET] {};

            \node [fit=(p0_1) (p2_1), MEMSET] {};

            \node [fit=(p0_2) (p2_2), MEMSET] {};

            \begin{scope}[on background layer]
                  \draw [SIM] (p2_2.north west) -- (p2_0.north west);
                  \draw [SIM] (p2_2.south west) -- (p2_0.south west);
            \end{scope}

            \matrix[MAT,
                  matrix of nodes,
                  column sep=\matColSep mm,
                  row sep=\matRowSep mm,
                  below=of static,
                  nodes={MEMBER, XS, anchor=center},
            ] (recon) {
                  \node (p0_0) {}; \&
                  \node (p0_1) {}; \&
                  \node[ADV] (p0_2) {}; \\
                  \node (p1_0) {}; \&
                  \node (p1_1) {}; \&
                  \node[ADV] (p1_2) {}; \\
                  \node[ADV] (p2_0) {}; \&
                  \node[ADV] (p2_1) {}; \&
                  \node[ADV] (p2_2) {}; \\
                  \node (p3_0) {}; \&
                  \node (p3_1) {}; \&
                  \node (p3_2) {};      \\
                  \node (p4_0) {}; \&
                  \node (p4_1) {}; \&
                  \node (p4_2) {};      \\
            };

            \node (M0) [fit=(p0_0) (p2_0), MEMSET] {};

            \node (M1) [fit=(p1_1) (p3_1), MEMSET] {};

            \node (M2) [fit=(p2_2) (p4_2), MEMSET] {};

            \draw [RANGE] (p0_2.north west) to[out=165,in=15] (p0_0.north east);
            \draw [RANGE] (p1_2.north west) to[out=165,in=15] (p1_0.north east);

            \matrix[MAT,
                  matrix of nodes,
                  column sep=\matColSep mm,
                  row sep=\matRowSep mm,
                  below=of da,
                  nodes={MEMBER, XS, anchor=center},
            ] (DAR) {
                  \node (p0_0) {}; \&
                  \node[ASLEEP] (p0_1) {}; \&
                  \node[ADV] (p0_2) {}; \\
                  \node[ASLEEP] (p1_0) {}; \&
                  \node (p1_1) {}; \&
                  \node[ADV] (p1_2) {}; \\
                  \node[ASLEEP] (p2_0) {}; \&
                  \node[ASLEEP] (p2_1) {}; \&
                  \node[ADV] (p2_2) {}; \\
                  \node[ASLEEP] (p3_0) {}; \&
                  \node[ASLEEP] (p3_1) {}; \&
                  \node (p3_2) {};      \\
                  \node[ASLEEP] (p4_0) {}; \&
                  \node[ASLEEP] (p4_1) {}; \&
                  \node (p4_2) {};      \\
            };

            \node (M0) [fit=(p0_0) (p2_0), MEMSET] {};
            \node (M1) [fit=(p1_1) (p3_1), MEMSET] {};
            \node (M2) [fit=(p2_2) (p4_2), MEMSET] {};

            \draw [RANGE] (p0_2.north west) to[out=165,in=15] (p0_0.north east);
            \draw [RANGE] (p1_2.north west) to[out=165,in=15] (p1_0.north east);

            \begin{scope}[on background layer]
                  \draw [SIM] (p2_2.north west) -- (p2_0.north west);
                  \draw [SIM] (p2_2.south west) -- (p2_0.south west);
            \end{scope}

            \coordinate (badge) at ([xshift=6mm, yshift=-3mm] current bounding box.east);
            \fill [black!85, shift={(badge)}]
            (90:2.7mm) -- (126:1.15mm) -- (162:2.7mm) -- (198:1.15mm) -- (234:2.7mm) --
            (270:1.15mm) -- (306:2.7mm) -- (342:1.15mm) -- (18:2.7mm) -- (54:1.15mm) -- cycle;
            \node[font=\bfseries\small, anchor=north] (dar) at ([yshift=-3mm] badge) {DAR};

            \node[draw=none, above=1mm of static] {Static availability};
            \node[draw=none, above=1mm of da] {Dynamic availability};

            \node[draw=none, left=5mm of static, align=center] {Non-reconfigurable};
            \node[draw=none, left=5mm of recon] {Reconfigurable};
      \end{tikzpicture}
      \caption[Models of reconfiguration and availability]{Models of reconfiguration and availability across three rounds.
            Here, ``\honpic'' depicts awake and honest nodes, ``\advpic'' are Byzantine nodes, and ``\dapic'' are asleep nodes.
            ``\longrangepic'' depicts long-range attacks, and ``\simpic'' depicts backward simulation.
      }
      \Description{Four small diagrams arranged in a two-by-two grid.
            The columns are labeled Static availability and Dynamic availability, and the rows are labeled Non-reconfigurable and Reconfigurable.
            In every diagram, each node is a row of markers and each column is one of three consecutive rounds, and a box is drawn around the nodes that form the membership set in that round.
            The two non-reconfigurable diagrams have three nodes, and their membership box covers the same three nodes in all three rounds.
            The two reconfigurable diagrams have five nodes, and their membership box shifts down by one node per round, so the top node leaves the membership set while a new node at the bottom joins it.
            In the two statically available diagrams every member is awake in every round, whereas in the two dynamically available diagrams several nodes are asleep in the earlier rounds.
            Byzantine markers appear in the later rounds of all four diagrams, and in the reconfigurable diagrams they also appear on nodes that have already left the membership set.
            In the reconfigurable diagrams, arcs curve backward from such a departed Byzantine node in the last round to the same node in the first round.
            In the dynamically available diagrams, double lines run backward from a Byzantine node in the last round through the earlier rounds in which that node was asleep.
            The bottom-right diagram combines asleep nodes, a shifting membership set, backward arcs, and backward double lines, and is marked with a star labeled DAR.}
      \label{fig:models}
\end{figure*}

We briefly describe the difficulties in solving consensus under the different combinations of availability and reconfiguration.
This will aid in our reasoning in the DAR model.
We categorize models along two orthogonal dimensions: availability (static vs.\ dynamic), and membership (non-reconfigurable vs.\ reconfigurable).
We label each model accordingly, such as ``Dynamic Availability and Reconfiguration (DAR)'' or ``Dynamic Availability and Non-Reconfiguration (Sleepy Model).''
We summarize the scenarios in \cref{fig:models}.

\paragraph*{Static Availability and Non-Reconfiguration}
These systems consider a fixed set of nodes where honest nodes are always available.
Classic protocols such as PBFT~\cite{CL99} operate in this setting.
Many protocols in this setting utilize digital signatures to bind nodes to the messages they send.
Once a node receives sufficiently many signed messages, the node can construct a quorum \emph{certificate}, a transferable proof that a particular event, or decision, has occurred with sufficient backing.

\paragraph*{Static Availability and Reconfiguration}
These systems largely operate in the same manner as the classic protocols above.
There is only a slight tweak in that the certificates produced throughout the protocol must be verified with respect to the corresponding membership set.
Therefore, verification usually starts from the last known membership set and \emph{iteratively} verifies every consensus decision and reconfiguration up to the most current output of the protocol.

Reconfiguration brings a subtlety known as the \emph{long-range attack}~\cite{DPP19}.
It may be unrealistic to expect an honest node that has since exited the system to maintain its honesty indefinitely.
Thus, it is most desirable to place fault bounds only on the latest set of members.
However, nodes that have exited the system still possess their cryptographic signing keys.
If enough of them are corrupted by the adversary, the adversary can generate certificates for an alternative sequence of past decisions, essentially \emph{rewriting history} in a way that appears valid to a newly joining node.

To mitigate this problem, Algorand~\cite{CM19} and Ouroboros~\cite{KRDO17,DGKR17,BGK+18} make use of forward-secure digital signatures~\cite{BM99,IR01,DGNW19,WTW+24} to prevent the adversary from forging signatures over past time periods after corrupting a node and obtaining its private state.

\paragraph*{Dynamic Availability and Non-Reconfiguration (Sleepy Model)}
These systems operate from the same fixed set of nodes throughout the entire execution, but nodes may ``go to sleep'' for an arbitrary amount of time.
As such, nodes that are awake must maintain liveness without the presence of all nodes.
The sleepy model, therefore, assumes synchrony because it is impossible to solve consensus in the sleepy model if the network can be partitioned~\cite{PS17,NTT20a,LR23} (see also \cref{sec:sleepy-sync-necessary}).

Unlike reconfiguration, all nodes are assumed members, so there is no need to distinguish nodes that left or joined as members since the start of execution.
Sleepy consensus protocols~\cite{PS17,MR22} assume an honest majority in the sense that awake and honest nodes outnumber adversarial nodes at any given time.
Adversarial nodes are assumed to always be awake.
Therefore, as participation grows, the adversary may corrupt more nodes.
This becomes increasingly problematic since the adversary may \emph{simulate} as if the newly corrupted nodes were already awake in the past.
Consider the execution shown in \cref{fig:models} in the sleepy model (top-right), where there is one awake node at time $0$, and three nodes awake at time $2$.
If we adapted techniques from the static availability setting, certificate sizes would be relative to the number of available nodes at the time of the event.

Thus, similarly to the long-range attack, the adversary may simulate alternative certificates and rewrite history in a way that appears valid to nodes waking up from their sleep---an attack known as \emph{backward simulation} or \emph{costless simulation}~\cite{DPS16,MMR23}.
Forward-secure digital signatures do not resolve the problem in this setting, since, as discussed above, asleep nodes cannot perform key evolution.
Recent works~\cite{DKT21,MR22,DNTT23,DSTZ24} solve this issue by completely ignoring past messages and only listening to the latest messages on what decisions to follow.

\paragraph*{Dynamic Availability and Reconfiguration (DAR)}
Naturally, establishing a secure protocol in the DAR model inherits all the difficulties from the two prior models: nodes may be asleep, and the membership set is not fixed.
Whenever a node is outdated, either because it has been asleep for some time or has recently joined the system, it must determine the latest state of the system using only the protocol logic and the initial membership set.
We call this process \emph{bootstrapping}.
After reasoning about the prior two models, one can recognize the main challenge for bootstrapping: \emph{current} active nodes alone dictate the state of the system (cf.\ Dynamic Availability and Non-Reconfiguration), but nodes bootstrapping into the system must verify \emph{past} reconfigurations to know who to listen to at present (cf.\ Static Availability and Reconfiguration).

Bitcoin, a longest-chain proof-of-work (PoW) protocol, has a simple bootstrapping procedure with minimum assumptions: a newly joining node only requires the protocol logic, the initial genesis block, and connection to at least one honest and up-to-date node.
Even if all other connections are to adversarial nodes, the newly joining node can distinguish the true chain by simply following the one with the most work.
Our bootstrapping gadgets in the DAR model (both plain and with sign-off) also require minimum common knowledge and trust assumptions that are conceptually similar to Bitcoin's bootstrapping procedure, though we show that a stronger variant of the honest majority condition is required for any consensus protocol to be secure in the plain DAR model.

\paragraph*{Organization.}

In \cref{model}, we outline basic aspects of our models that are used throughout the paper.
In \cref{dar-recon}, we formalize reconfigurable consensus.
In \cref{dar-gadget}, we present our bootstrapping gadget for constructing secure atomic broadcast protocols in the DAR model under the \sr condition,
and prove that \srfulllight is necessary in \cref{dar-srhm-necessary}.
We consider signed membership changes (DAR with sign-off model) and give a bootstrapping gadget with a simple and efficient good-case path in \cref{darso}.
We discuss further related work in \cref{related}.

\section{Model and Preliminaries}
\label{model}

\subsection{Model}

We assume a public key infrastructure (PKI), where each node's public key is known to all others.

\paragraph*{Synchrony.}

\label{sec:sleepy-sync-necessary}

We consider a lock-step synchronous system in which nodes proceed in globally synchronized discrete \emph{rounds} and are assumed to have synchronized clocks.
Messages seen by \emph{any} honest (and awake) node by the start of a round are seen by \emph{all} honest (and awake) nodes by the end of the round.
Subject to this constraint, the adversary can selectively reveal messages it sends to honest nodes.
Note that all prior works on dynamic availability assume synchrony~\cite{PS17,NTT20a,MR22,MMR23,DSTZ24,DNTT23,DPS16,BGK+18}.
It is easy to see that consensus with dynamic availability is impossible with partial synchrony or asynchrony~\cite{LR24b}.
We now give a proof sketch.
In a partially synchronous or asynchronous network, the adversary can partition awake and honest nodes into two disjoint sets and delay all messages between the two sets.
To each set, the other set appears to be asleep.
Due to dynamic availability and liveness, both sets must eventually decide, even though they never communicate with each other.
The decisions will conflict, violating safety.

\paragraph*{Adversary.}

The adversary is a probabilistic polynomial-time (PPT) algorithm that can fully adaptively \emph{corrupt} nodes and put nodes to \emph{sleep}.
Corrupted nodes (also called \emph{adversarial} or \emph{Byzantine}) may deviate from the protocol arbitrarily in a manner coordinated by the adversary.
We denote the corrupted nodes at a round $\rnd$ as $\adv_{\rnd}$.
Once a node is corrupted, the node will never become honest again, so $\adv_{\rnd} \subseteq \adv_{\rnd + 1}$.
Uncorrupted nodes (called \emph{honest}) follow the protocol when they are not asleep (as discussed below).

\paragraph*{Sleepiness.}

Besides corruption, the adversary can fully adaptively put nodes to \emph{sleep} at each round.
\emph{Asleep} nodes do not receive or send messages and do not execute the protocol.
When an asleep node wakes up, it enters a state of \emph{booting up} and receives any critical messages required to be up-to-date.
During this time, the node initiates a \emph{bootstrapping} process, after which it becomes fully \emph{awake}.
The process mirrors Bitcoin, where nodes must receive all critical block data to reconstruct the history.
In practice, this bootstrapping is interactive and may span many rounds, with nodes querying neighbors for critical messages on demand and potentially re-bootstrapping later (e.g., when discovering a longer chain in Bitcoin).
For simplicity, we model this interactive retrieval as instantaneous, as if sleeping nodes had buffered all necessary messages---though in practice, no such buffer is needed since nodes retrieve messages on demand upon waking.
An awake and honest node executes the specified protocol.
We denote awake and honest nodes at a round $\rnd$ as $\hon_{\rnd}$.
Adversarial nodes are always considered awake and cannot be put to sleep.

\subsection{Atomic Broadcast}
\label{model-definitions}

\paragraph*{Log.}

We define a \emph{log} as a sequence of values $x_i$, denoted as $\chain = [x_0, x_1, x_2, ...]$.
We say two logs are \emph{compatible} if one is a prefix of the other.
Otherwise, we say they are \emph{conflicting}.
At every round, each node has an output log.

\paragraph*{Atomic Broadcast.}

An atomic broadcast protocol~\cite{CASV95} allows nodes to agree on a log.
Nodes may input requests into the protocol at each round.
Logs that are output across honest nodes and across time satisfy the following properties:
\begin{itemize}
      \item \emph{Safety:}
            If two honest nodes (possibly the same node) at two points in time (possibly the same time) decide logs $\chain$ and $\chain'$, then $\chain$ and $\chain'$ are compatible.

      \item \emph{Liveness:}
            If an awake and honest node inputs a value $x$, then eventually all awake and honest nodes decide a log containing $x$.

      \item \emph{Validity:}
            If an honest log contains the value $x$, then $x$ is input by some node.
\end{itemize}

\subsection{Forward-Secure Signatures}
We make use of a forward-secure signature scheme for signing messages in our protocols.
\emph{Forward security} ensures unforgeability of messages in the past---prior to key exposure---such that only messages after key exposure are vulnerable.
Bellare and Miner's~\cite{BM99} forward-secure signature construction consists of a single public key $\pk$ and an initial signing key denoted $\sk_0$.
They consider $T$ time periods in which the signature scheme is used.
The signer uses a different signing key in each period: $\sk_1$ in period $1$, $\sk_2$ in period 2, and so on.
At the beginning of period $i$, $\sk_i$ is derived as a one-way function on $\sk_{i-1}$.
This is the process of \emph{evolving} the key, and, once complete, the signer deletes the previous key $\sk_{i-1}$.
Thus, an adversary that corrupts the signer during period $i$ cannot forge messages in prior periods.
Let $\kes$ be a key-evolving signature scheme parametrized by security parameter $\lambda$ and the total number of time periods $T$.
We define $\kes$ with the following four algorithms taken from Bellare and Miner~\cite{BM99} for a signer $v$:
\begin{itemize}
      \item \textbf{Key Generation:} $(v.\pk,\sk_0)\gets\kes.\Call{Gen}{ }$.
            Signer $v$ runs the key generation algorithm to generate a public verification key $\pk$ and an initial secret signing key $\sk_0$.

      \item \textbf{Key Update:} $\sk_{i + 1} \gets \kes.\Call{Update}{\sk_{i}}$.
            Signer $v$ updates its secret key $\sk_i$ to $\sk_{i+1}$ for the next time period $i+1$.
            We may also use $\sk_{i'} \gets \kes.\Call{Update}{\sk_{i},i'}$ to repeatedly apply the update to any time period $i' > i$.

      \item \textbf{Sign:} $\signedt{m}{v}{i} \gets \kes.\Call{Sign}{\sk_i, m}$.
            Signer $v$ uses the current signing key $\sk_i$ and a message $m$ to compute signature $\signedt{m}{v}{i}$.
            For convenience, we assume the signature $\signedt{m}{v}{i}$ carries the signer's identity $v$ and the time period $i$ in which the signature is produced.

      \item \textbf{Verification:} $b \gets \kes.\Call{Verify}{\signedt{m}{v}{i}}$.
            Anyone may verify the signature $\signedt{m}{v}{i}$ over message $m$ for time period $i$ under the public key $v.\pk$ by running the verification algorithm, which outputs $1$ if the signature is valid and $0$ otherwise.
\end{itemize}

As an example, we can use the Pixel forward-secure signature~\cite{DGNW19}, which supports key update, signing, and verification in $O(1)$ cryptographic operations (i.e., group exponentiations and pairings).
Both the signature size and public key size are constant.
The secret key size is $O(\log^2 T)$ group elements, where $T$ is an upper bound on the number of time periods for which the scheme is used.
For instance, if each time period lasts one second, then $T=2^{32}-1$ periods would be $136$ years, giving a secret key of $43$KB in size~\cite{DGNW19}.

\section{Bootstrapping in Plain DAR}
\label{dar}

This section presents our detailed model for reconfiguration (\cref{dar-recon}), our atomic broadcast protocol (\cref{dar-gadget}), and a matching impossibility result (\cref{dar-srhm-necessary}) in the plain DAR model.
We make no use of extra features through additional external mechanisms or node capabilities to solve the bootstrapping problem.
In contrast, in \cref{darso}, we will study a variant of the model where exiting nodes must authorize their withdrawal, as is the case in PoS systems.

\subsection{Modeling Reconfiguration}
\label{dar-recon}

The initial membership set $\genesis$, a finite set of nodes, is common knowledge.
Subsequent membership sets $\mem_{\rnd}$, for rounds $\rnd > 0$, are not known a priori to all nodes.
Note that an honest node knows whether it is awake, but not whether it is a current member.
We assume a fixed membership size, i.e., $|\genesis| = |\mem_{\rnd}|$ for all rounds $\rnd \geq 0$.

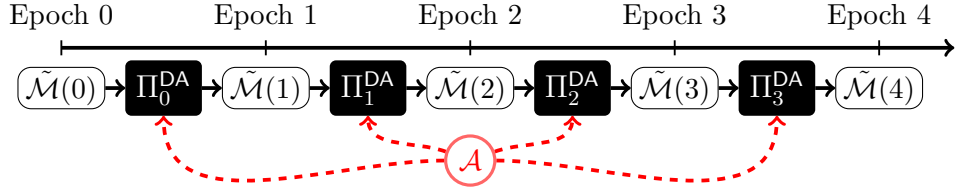
\begin{figure*}[tb]
    \centering
    \begin{tikzpicture}
        \matrix[MAT,
            matrix,
            column sep=1.55cm,
            row sep=1.25mm,
            at={(0,0)},
            nodes={anchor=center},
        ] (nodes) {
            \node (tick0) [opacity=0] {};                         \&
            \node (tick1) [opacity=0] {};                         \&
            \node (tick2) [opacity=0] {};                         \&
            \node (tick3) [opacity=0] {};                         \&
            \node (tick4) [opacity=0] {};                   \\

            \node (M0) [MEMSET] {$\emem[0]$};     \&
            \node (M1) [MEMSET] {$\emem[1]$};     \&
            \node (M2) [MEMSET] {$\emem[2]$};     \&
            \node (M3) [MEMSET] {$\emem[3]$};     \&
            \node (M4) [MEMSET] {$\emem[4]$}; \\
        };

        \node (P0) [BLACKBOX] at ($(M0)!0.5!(M1)$) {$\eda[0]$};
        \draw [RECON] (M0) -- (P0) {};
        \draw [RECON] (P0) -- (M1) {};

        \node (P1) [BLACKBOX] at ($(M1)!0.5!(M2)$) {$\eda[1]$};
        \draw [RECON] (M1) -- (P1) {};
        \draw [RECON] (P1) -- (M2) {};

        \node (P2) [BLACKBOX] at ($(M2)!0.5!(M3)$) {$\eda[2]$};
        \draw [RECON] (M2) -- (P2) {};
        \draw [RECON] (P2) -- (M3) {};

        \node (P3) [BLACKBOX] at ($(M3)!0.5!(M4)$) {$\eda[3]$};
        \draw [RECON] (M3) -- (P3) {};
        \draw [RECON] (P3) -- (M4) {};

        \node (A) [MEMBER, ADV] at ($(tick2) + (0, -1.5cm)$) {$\adv$};
        \draw[use as bounding box, opacity=0] (current bounding box.north west) rectangle ($(tick4.east |- A.south)$);
        \draw [ADVRECON] (A) to[out=180,in=-90] (P0.south);
        \draw [ADVRECON] (A) to[out=160,in=-90] (P1.south);
        \draw [ADVRECON] (A) to[out=20,in=-90] (P2.south);
        \draw [ADVRECON] (A) to[out=0,in=-90] (P3.south);

        \def\numTicks{5}
        \coordinate (tick5) at ($(tick4) + (1cm, 0)$);
        \draw [->, line width=0.5mm] (tick0.center) -- (tick5) {};
        \pgfmathtruncatemacro{\lastTick}{\numTicks - 1}
        \foreach \i in {0,...,\lastTick} {
                \draw [-, line width=0.25mm] ($(tick\i.center) + (0, -1mm)$) -- ($(tick\i.center) + (0, 1mm)$) node[above] {Epoch $\i$};
            }
    \end{tikzpicture}
    \caption{
        Overview of reconfiguration every epoch. ``\advreconpic'' indicates the adversary's membership input into $\eda$.
    }
    \Description{A left-to-right timeline with five tick marks labeled Epoch 0 through Epoch 4.
        Above each tick sits a box denoting the membership set of that epoch.
        Between every pair of consecutive membership sets sits a black box denoting the protocol instance run during that epoch, with an arrow from the earlier membership set into the box and a further arrow from the box into the later membership set, so the five membership sets and four protocol instances form a single chain.
        Below the timeline sits a marker for the adversary, from which four curved arrows rise, one into the underside of each of the four protocol instances.}
    \label{fig:recon}
\end{figure*}

\paragraph*{Epochs.}
Time is divided into fixed-length epochs of $\nrounds$ rounds, a parameter chosen beforehand that is common knowledge.
Membership is fixed during each epoch and may change across epochs.
Denote the \emph{decided} membership for epoch $\epoch$ as $\emem[\epoch]$.
Then, we have $\emem[\epoch] = \mem_{\epoch\nrounds} = \mem_{\epoch\nrounds + 1} = ... = \mem_{\epoch\nrounds + \nrounds - 1}$.

Let $\eda$ be the instance of a dynamically available, non-re\-con\-fig\-ur\-able atomic broadcast protocol $\da$ associated with epoch $\epoch$, which is run by nodes in $\emem$.
At each epoch $\epoch \geq 0$, the adversary may select a new set of nodes $\memvar$ and input it into the atomic broadcast protocol $\eda$ through its own input channel.
At the end of the epoch, awake and honest nodes can set $\emem[\epoch + 1]$ as the latest decided membership set in their log and use it as the next epoch's membership.
We illustrate the process in \cref{fig:recon}.
An external mechanism that instead announces each membership to all nodes would amount to social consensus: an outdated node could learn the current membership by simply waiting for the next announcement.
Instead, letting the adversary propose membership sets captures that many possible sets can exist, and a node not participating in $\eda$ cannot tell which one took effect.
Agreement through $\eda$ guarantees that the awake and honest nodes in $\emem$ settle on a single membership set $\emem[\epoch + 1]$ and are the only nodes guaranteed to learn of it.

The liveness of $\eda$ ensures that $\emem[\epoch + 1]$ is \emph{eventually} included in the logs of all awake and honest nodes.
However, liveness alone does not guarantee that $\emem[\epoch + 1]$ is included \emph{by all awake and honest nodes by the end of epoch $\epoch$}.
Specifically, the concern is that, at the end of epoch $\epoch$, some honest nodes have $\emem[\epoch + 1]$ in their output log, while other honest nodes do not.
Thus, we require an additional property of $\eda$ that ensures the membership set $\emem[\epoch + 1]$ is in the decided logs of all awake and honest nodes at the end of epoch $\epoch$.
Such a property is easily attainable for common atomic broadcast implementations with simple transformations, albeit with a slight reduction in performance.
For instance, given a dynamically available protocol with a worst-case latency of $k$ rounds, a possible transformation is to require ``empty proposals'' during the last $k$ rounds of the epoch.
The last $k$ rounds do not contribute new inputs, decreasing the protocol's throughput.
However, it provides adequate time for all awake and honest nodes to settle on a common output by the end of the epoch.
Nodes then wait to apply any proposed membership set until the end of the epoch.
Under this model, we define the standard honest majority of a schedule (see \cref{def:schedule}) in \cref{def:hon-maj}, which we show in \cref{dar-srhm-necessary} is insufficient to guarantee safety and liveness of an atomic broadcast protocol in the DAR model.
Note that honest majority is the tight condition under which consensus is feasible in each of the two settings that the DAR model combines: static availability with reconfiguration~\cite{KRDO17,DGKR17} and dynamic availability without reconfiguration (i.e., the sleepy model~\cite{PS17,MR22}).
The insufficiency of an honest majority in the DAR model thus formalizes the sense in which the combination is harder than its parts.

\begin{definition}[Schedule]
    \label{def:schedule}
    A schedule $\schd$ defines the sequences of sets $\adv_\rnd$, $\hon_\rnd$, and $\mem_\rnd$ of adversarial nodes, awake and honest nodes, and membership at every round $\rnd \geq 0$.
    Note that the corruption, sleep, and membership schedule is chosen by the adversary.
\end{definition}
\begin{definition}[Honest Majority (HM)]
    \label{def:hon-maj}
    A schedule satisfies \emph{honest majority} if for every round $\rnd \geq 0$,
    \[
        |\mem_{\rnd} \cap \adv_{\rnd}| < |\mem_{\rnd} \cap \hon_{\rnd}|.
    \]
\end{definition}

\paragraph*{Simulation.}

To capture what is sufficient and necessary for solving consensus in the DAR model, we pivot to the notion of \emph{simulation}.
While asleep, a node does not perform protocol actions: it neither sends nor receives messages, nor evolves its keys.
If such a node later becomes corrupted, the adversary can retroactively \emph{simulate} its behavior, making it appear as though it had actively participated in past rounds despite being asleep.
We call such nodes \emph{simulatable}.
Note that the node is only simulatable up to the last time it was awake and honest, since it may have performed a key evolution at that point, preventing the adversary from signing messages for earlier rounds.
\Cref{def:simulatable} gives the precise definition of simulatable nodes.

\begin{definition}[Simulatable Nodes]
    \label{def:simulatable}
    Denote the set of nodes that have been awake and honest in any round during the range of rounds $[\srnd, \rnd]$ as
    \[
        \hons{\srnd}{\rnd} \triangleq \bigcup_{\rnd' \in [\srnd, \rnd]} \hon_{\rnd'}.
    \]
    Then, denote the set of nodes that are simulatable at round $\srnd$ by the adversary at round $\rnd$ as
    \[
        \sims{\srnd}{\rnd} \triangleq \adv_{\rnd} \setminus \hons{\srnd}{\rnd}.
    \]
\end{definition}

Using \cref{def:simulatable}, we introduce the \emph{simulation-resistant honest majority} (\sr) in \cref{def:srhm}, which strengthens the standard honest majority (\hm, \cref{def:hon-maj}).
In \cref{dar-srhm-necessary}, we prove that the standard honest majority over every round $\rnd$ is insufficient for consensus.

\begin{definition}[Simulation-Resistant Honest Majority (\sr)]
    \label{def:srhm}
    A schedule (\cref{def:schedule}) satisfies \emph{simulation-resistant honest majority} if for all rounds $\rnd \geq 0$ and $\srnd \leq \rnd$, the adversarial and simulatable nodes at round $\srnd$ are fewer than the awake and honest nodes since round $\srnd$, i.e.,\
    \[
        \left|\mem_{\srnd} \cap \sims{\srnd}{\rnd}\right| < \left|\mem_{\srnd} \cap \hons{\srnd}{\rnd}\right|.
    \]
\end{definition}

\subsection{Bootstrapping Gadget in Plain DAR with Simulation-Resistant Honest Majority}
\label{dar-gadget}

In this section, we present an atomic broadcast protocol in the DAR model under \sr by layering a bootstrapping gadget atop the dynamically available, non-reconfigurable protocol $\da$.
The bootstrapping gadget enables outdated nodes to learn the current membership set and join the current epoch's instance of $\da$.
At the end of each epoch, the gadget has awake nodes produce certification votes for the latest decided log, which also behaves as a vote for every prefix of the log.
Booting nodes collect these votes upon awakening.
They follow the log with the most votes from their last known membership set at each epoch until they reach the latest epoch.
Although the decided log at each epoch may not have an absolute quorum of votes due to dynamic availability, we prove it safe to follow the log with the ``heaviest votes'' under \sr, since, for every round, the number of awake and honest members is greater than the number of simulatable members.
The bootstrapping gadget is detailed in \cref{alg:dar-gadget}.

\begin{algorithm*}[tb]
    \caption{Reconfigurable atomic broadcast for node $v$}
    \label{alg:dar-gadget}
    \begin{algorithmic}[1]
        \During{epoch $\epoch$, \textbf{if} awake}%
            \label{alg:dar-gadget:during-epoch}

            \State Execute $\da$ among members in $\memvar_{\epoch}$ and store output in $\chain$%
                \label{alg:dar-gadget:static}
            \At{the last round of epoch $\epoch$}

                \State $\signedt{\chain}{v}{\epoch} \gets \kes.\Call{Sign}{\sk_{\epoch},\chain}$%
                    \Comment{Sign with forward-secure signature}%
                    \label{alg:dar-gadget:sign}

                \State Send $\logmsg{\signedt{\chain}{v}{\epoch}}$ to all%
                    \Comment{Send out vote for latest log}%
                    \label{alg:dar-gadget:vote}

                \State $\memvar_{\epoch + 1} \gets \getmem{\chain}$%
                    \Comment{Get membership set for next epoch from log}%
                    \label{alg:dar-gadget:extract-mem}

                \State $\sk_{\epoch + 1} \gets \kes.\Call{Update}{\sk_\epoch}$%
                    \Comment{Evolve secret key to latest epoch}%
                    \label{alg:dar-gadget:evolve}
            \EndAt
        \EndDuring

        \Upon{boot up during epoch $\epoch$}%
            \label{alg:dar-gadget:on-boot}
            \If{first time booting up}%
                \label{alg:dar-gadget:first-boot}
                \State $(\pk, \sk_0) \gets \kes.\Call{Gen}{ }$%
                    \Comment{Generate keys}%
                    \label{alg:dar-gadget:keys}

                \State $\memvar_0 \gets \genesis$%
                    \Comment{Initialize as initial membership set}%
                    \label{alg:dar-gadget:init}
                \State $\ell \gets 0$
            \Else
                \State $\ell \gets$ last epoch $v$ was awake%
                    \label{alg:dar-gadget:last-awake}
            \EndIf

            \State Receive $\logmsgtype$ messages from network with valid forward-secure signatures%
                \label{alg:dar-gadget:get-votes}

            \For{$\epoch' \gets \ell, \ell + 1, ..., \epoch - 1$}
                \State $\votes \gets$ oldest $\logmsgtype$ messages at or after epoch $\epoch'$ from members in $\memvar_{\epoch'}$%
                    \label{alg:dar-gadget:proposed-logs}

                \State $\chain_{\epoch'} \gets $ most voted log up to epoch $\epoch'$ in $\votes$%
                    \label{alg:dar-gadget:get-most-voted-log}

                \State $\memvar_{\epoch' + 1} \gets \getmem{\chain_{\epoch'}}$%
                    \Comment{Get membership set for next epoch from log}%
                    \label{alg:dar-gadget:next}
            \EndFor

            \State $\sk_{\epoch} \gets \kes.\Call{Update}{\sk_{\ell}, \epoch}$
        \EndUpon
    \end{algorithmic}
\end{algorithm*}

\paragraph*{Awake Participation.}
Awake nodes send inputs to $\eda$ during epoch $\epoch$ and output the decided log (\cref{alg:dar-gadget:static}).
At the end of the epoch, nodes produce a forward-secure signature over the log output $\chain$ to send out as a \emph{vote} over the decided log (\cref{alg:dar-gadget:sign,alg:dar-gadget:vote}).
Then, nodes obtain the next membership set from $\chain$, denoted as $\memvar_{\epoch + 1}$ for membership at epoch $\epoch + 1$ (\cref{alg:dar-gadget:extract-mem}).
Lastly, awake nodes evolve their secret keys to the next epoch (\cref{alg:dar-gadget:evolve}).

\paragraph*{Bootstrapping.}
The bootstrapping process starts at \cref{alg:dar-gadget:on-boot} when a node $v$ boots up during an epoch $\epoch$.
If this is the first time $v$ is awake, it initializes its keys for the forward-secure signature scheme $\kes$ and sets its last known membership set to the initial membership set $\genesis$.
The booting node $v$ then receives all $\logmsgtype$ messages and verifies their signatures (\cref{alg:dar-gadget:get-votes}).
The bootstrapping process proceeds from $v$'s last known membership set $\memvar_{\ell}$ at epoch $\ell$.
For every epoch $\epoch' \in [\ell, \epoch)$, $v$ collects the oldest votes at or after epoch $\epoch'$ for logs from nodes in $\memvar_{\epoch'}$ as the set $\votes$ (\cref{alg:dar-gadget:proposed-logs}).
We are interested in the oldest vote since an honest node may be corrupted after issuing a vote, and we want to ensure that we capture all honest votes concerning epoch $\epoch'$.
At \cref{alg:dar-gadget:get-most-voted-log}, $v$ tallies the number of votes for logs up to epoch $\epoch'$ and selects the log with the most votes as $\chain_{\epoch'}$.
Then, $v$ extracts the next membership set $\memvar_{\epoch' + 1}$ from $\chain_{\epoch'}$ (\cref{alg:dar-gadget:next}).
After iterating through every epoch, $v$ has identified the latest membership set and joins the atomic broadcast protocol $\eda$.

\paragraph*{Efficiency Overhead.}
Beyond its participation in $\da$, each awake node additionally sends a $\logmsgtype$ message to all nodes at the end of every epoch.
Any node booting up at epoch $\epoch$ verifies the $\logmsgtype$ messages for each missed epoch since it was last awake.

\begin{theorem}
    \label{thm:dar-gadget}
    Under \srfull, the protocol shown in \cref{alg:dar-gadget} implements atomic broadcast in the plain DAR model.
\end{theorem}

\deferredproof{proof}{thm:dar-gadget}

\begin{defer}{proof}
    \begin{proof}[Proof of \cref{thm:dar-gadget}]
        For every epoch $\epoch$, since the membership is fixed for the entire epoch, the protocol is safe, live, and valid for awake and honest nodes due to the safety, liveness, and validity of $\eda$.
        It only remains to show that if an honest node $v$ booting up during epoch $\epoch$ participates, it does so with the same membership set $\emem[\epoch]$ as the awake and honest nodes in that epoch.

        We will inductively show that \crefrange{alg:dar-gadget:on-boot}{alg:dar-gadget:next} identify the decided membership sets from $0$ to $\epoch$.
        Let $\rnd$ be the round in epoch $\epoch$ in which node $v$ is booting up.
        This is the same round and epoch as current awake nodes due to synchronized clocks.
        For the base case, we have that the initial membership set $\genesis$ is common knowledge.
        For the inductive step, assume that $v$ has selected the decided membership set $\emem[\epoch']$ for epoch $\epoch' < \epoch$.
        We now show that $v$ selects $\emem[\epoch' + 1]$ as the confirmed membership set for epoch $\epoch' + 1$ at \cref{alg:dar-gadget:next}.

        Consider a false log $\bad{\chain}$ that is conflicting with the decided log $\echain[\epoch']$.
        Let $c$ and $\bad{c}$ be the total count of votes for $\echain[\epoch']$ and $\bad{\chain}$, respectively, in $\votes$ (\cref{alg:dar-gadget:proposed-logs}).
        Let $\srnd$ be the last round of epoch $\epoch'$.
        By the induction hypothesis, all observed votes from round $\srnd$ are from members in $\emem[\epoch']$.
        Thus, we must have that $c \geq |\mem_{\srnd} \cap \hons{\srnd}{\rnd}|$, because every honest member in $\mem_{\srnd}$ that was awake at some round at or after $\srnd$ signed and sent out a vote for a log that extends $\echain[\epoch']$.
        We also have that $\bad{c} \leq |\mem_{\srnd} \cap (\adv_{\srnd} \cup \sims{\srnd}{\rnd})|$ since, by the definition of the forward-secure signature scheme, only adversarial and simulatable nodes at round $\srnd$ have the secret keys available to sign a vote for a false log.
        Since $\adv_{\srnd} \subseteq \sims{\srnd}{\rnd}$, by \srfull, we have that
        \[
            \bad{c} \leq |\mem_{\srnd} \cap \sims{\srnd}{\rnd}| < |\mem_{\srnd} \cap \hons{\srnd}{\rnd}| \leq c.
        \]
        Thus, we must have that the most voted log for epoch $\epoch'$ is $\echain[\epoch']$, which $v$ uses to retrieve the next decided membership set $\emem[\epoch' + 1]$.
        By induction, we have that node $v$ identifies the latest membership set $\memvar_{\epoch} = \emem[\epoch]$.
    \end{proof}
\end{defer}

\subsection{Necessity of Simulation-Resistant Honest Majority}
\label{dar-srhm-necessary}
We now show that \sr is necessary for any atomic broadcast protocol to be safe, live, and valid in the DAR model.
Importantly, we restrict our attention to \emph{identity-agnostic} conditions, meaning that relabeling nodes in a schedule does not change whether the condition holds.
We prove that no \emph{identity-agnostic} condition that fails to imply \sr can suffice for an atomic broadcast protocol.
Note that both the \hm and \sr conditions are identity-agnostic.

\begin{definition}[Identity-Permuted Schedule]
    \label{def:permuted-schedule}
    For a schedule $\schd$ and a permutation $\nodemap$ of node identities, define the \emph{identity-permuted schedule} $\nodemap(\schd)$ to be the schedule in which node $\nodemap(v)$ inherits the full local view and state (i.e., corruption, honesty, and membership) of node $v$ across all rounds in $\schd$.
    We say two schedules $\schd$ and $\schd'$ are identity-permuted schedules if there exists a permutation $\nodemap$ such that $\schd' = \nodemap(\schd)$.
\end{definition}

\begin{definition}[Identity-Agnostic Condition]
    \label{def:identity-agnostic}
    A condition $\cond$ over schedules is said to be \emph{identity-agnostic} if for all schedules $\schd$, whenever $\cond$ holds for $\schd$, $\cond$ also holds for all permuted schedules of $\schd$, i.e., for all schedules $\schd$ and permutations $\phi$, we have that
    \[
        \cond(\schd) \iff \cond(\phi(\schd)).
    \]
\end{definition}

\begin{definition}[Execution]
    We define an \emph{execution} $\exec$ as a sequence of messages and states of all nodes under a schedule $\schd$ during execution of a protocol $\protocol$.
\end{definition}

\begin{theorem}
    \label{thm:srhm-necessary}
    No atomic broadcast protocol can be safe, live, and valid in the plain DAR model under an identity-agnostic condition (\cref{def:identity-agnostic}) that does not imply \srfull.
\end{theorem}

\deferredproof{proof}{thm:srhm-necessary}

\begin{defer}{proof}
    \begin{proof}[Proof of \cref{thm:srhm-necessary}]
        \begin{figure*}[tb]
    \centering
    \begin{tikzpicture}[node distance=2mm,
            WITHDRAWN/.style={draw=none, fill=black, opacity=0.15, text opacity=0},
            FADEBOUND/.style={MEMBER, draw=none, fill=white, opacity=0.75, text opacity=0},
        ]
        \def\centerX{0}
        \def\centerY{0}
        \def\tickSize{0.25}

        \def\offsetTitleY{0mm}

        \def\offsetMEMY{2mm}
        \def\offsetMEMX{6mm}
        \def\offsetRounds{4mm}
        \edef\offsetNodes{\iffull5mm\else4mm\fi}

        \def\offsetBootLOGX{4mm}

        \def\offsetExecY{15mm}

        \coordinate (center) at (\centerX, \centerY);

        \matrix[MAT, column sep=\offsetNodes, row sep=\offsetRounds, nodes={MEMBER, anchor=center}, below left=\offsetMEMY and \offsetMEMX of center.south] (execution1) {
            \node (Q1_0) {$Q_1$};           \&
            \node (Q2_0) [ASLEEP] {$Q_2$}; \&
            \node (Q3_0) [ASLEEP] {$Q_3$}; \&
            \node (Q4_0) [ASLEEP, ADV] {$Q_4$}; \\

            \node (Q1_1) {$Q_1$};          \&
            \node (Q2_1) [ASLEEP, ADV] {$Q_2$};    \&
            \node (Q3_1) {$Q_3$}; \&
            \node (Q4_1) [ASLEEP, ADV] {$Q_4$}; \\
        };

        \node (M0) [MEMSET, fit=(Q1_0) (Q2_0)] {};
        \begin{scope}[on background layer]
            \node (M1_0) [MEMSET, inner xsep=0pt, fit={($(Q1_1.center) + (-0.5mm, 0)$) ($(Q2_1.center) + (0.5mm, 0)$) (Q1_1.north) (Q1_1.south)}] {};
            \node [FADEBOUND] at (Q1_1) {$Q_1$};
            \node [FADEBOUND] at (Q2_1) {$Q_2$};
        \end{scope}

        \begin{scope}
            \clip ($(Q1_1.center) + (-5mm, -5mm)$) rectangle ($(Q1_1.center) + (-0.5mm, 5mm)$);
            \node [MEMBER, WITHDRAWN] at (Q1_1) {$Q_1$};
        \end{scope}
        \begin{scope}
            \clip ($(Q2_1.center) + (0.5mm, -5mm)$) rectangle ($(Q2_1.center) + (5mm, 5mm)$);
            \node [MEMBER, WITHDRAWN] at (Q2_1) {$Q_2$};
        \end{scope}
        \node (M1_1) [MEMSET, fit=(Q3_1)] {};

        \node (msg1) [CALLOUT, BLUEMSG, below=of Q1_1] {$\msgs_{1}$};

        \node[above=\offsetTitleY of execution1] {\textbf{Execution~$\exec_1$}};

        \matrix[MAT,column sep=\offsetNodes, row sep=\offsetRounds, nodes={MEMBER, anchor=center}, below right=\offsetMEMY and \offsetMEMX of center.south] (execution2) {
            \node (Q1_0) [ASLEEP] {$Q_1$};  \&
            \node (Q2_0) {$Q_2$}; \&
            \node (Q3_0) [ASLEEP, ADV] {$Q_3$}; \&
            \node (Q4_0) [ASLEEP] {$Q_4$}; \\

            \node (Q1_1) [ASLEEP, ADV] {$Q_1$};     \&
            \node (Q2_1) {$Q_2$}; \&
            \node (Q3_1) [ASLEEP, ADV] {$Q_3$}; \&
            \node (Q4_1)  {$Q_4$};         \\
        };

        \node (M0) [MEMSET, fit=(Q1_0) (Q2_0)] {};
        \begin{scope}[on background layer]
            \node (M1_0) [MEMSET, inner xsep=0pt, fit={($(Q1_1.center) + (-0.5mm, 0)$) ($(Q2_1.center) + (0.5mm, 0)$) (Q1_1.north) (Q1_1.south)}] {};
            \node [FADEBOUND] at (Q1_1) {$Q_1$};
            \node [FADEBOUND] at (Q2_1) {$Q_2$};
        \end{scope}

        \begin{scope}
            \clip ($(Q1_1.center) + (-5mm, -5mm)$) rectangle ($(Q1_1.center) + (-0.5mm, 5mm)$);
            \node [MEMBER, WITHDRAWN] at (Q1_1) {$Q_1$};
        \end{scope}
        \begin{scope}
            \clip ($(Q2_1.center) + (0.5mm, -5mm)$) rectangle ($(Q2_1.center) + (5mm, 5mm)$);
            \node [MEMBER, WITHDRAWN] at (Q2_1) {$Q_2$};
        \end{scope}
        \node (M1_1) [MEMSET, fit=(Q4_1)] {};

        \node (msg2) [CALLOUT, ORANGEMSG, below=of Q2_1] {$\msgs_{2}$};

        \node[above=\offsetTitleY of execution2] (title2) {\textbf{Execution~$\exec_2$}};

        \path let \p1 = (title2.south) in coordinate (top) at (\centerX, \y1);

        \path let \p1 = (Q1_0.center) in coordinate (t') at (\centerX, \y1);
        \draw[-] ($(t') + (-\tickSize, 0)$) -- ($(t') + (\tickSize, 0)$) node[midway, left=2mm] {$\srnd$};

        \path let \p1 = (Q1_1.center) in coordinate (t) at (\centerX, \y1);
        \draw[-] ($(t) + (-\tickSize, 0)$) -- ($(t) + (\tickSize, 0)$) node[midway, left=2mm] {$\rnd$};

        \path let \p1 = (msg1.south) in coordinate (title_anchor) at (\centerX, \y1);
        \draw[-, loosely dotted, thick] (top) -- (t') {};
        \draw[-, dashed, thick] (t') -- (t) {};
        \draw[->, thick] (t) -- (title_anchor) node (timeline) {};%

        \matrix[MAT, column sep=\offsetNodes, row sep=\offsetRounds, nodes={MEMBER, anchor=center}, below=\offsetExecY of execution1] (execution1') {
            \node (Q1_0) {$Q_1$};           \&
            \node (Q2_0) [ASLEEP] {$Q_2$}; \&
            \node (Q3_0) [ASLEEP] {$Q_3$}; \&
            \node (Q4_0) [ADV] {$Q_4$}; \\

            \node (Q1_1) {$Q_1$};          \&
            \node (Q2_1) [ADV] {$Q_2$};    \&
            \node (Q3_1) {$Q_3$}; \&
            \node (Q4_1) [ADV] {$Q_4$}; \\
        };

        \node (M0) [MEMSET, fit=(Q1_0) (Q2_0)] {};
        \begin{scope}[on background layer]
            \node (M1_0) [MEMSET, inner xsep=0pt, fit={($(Q1_1.center) + (-0.5mm, 0)$) ($(Q2_1.center) + (0.5mm, 0)$) (Q1_1.north) (Q1_1.south)}] {};
            \node [FADEBOUND] at (Q1_1) {$Q_1$};
            \node [FADEBOUND] at (Q2_1) {$Q_2$};
        \end{scope}

        \begin{scope}
            \clip ($(Q1_1.center) + (-5mm, -5mm)$) rectangle ($(Q1_1.center) + (-0.5mm, 5mm)$);
            \node [MEMBER, WITHDRAWN] at (Q1_1) {$Q_1$};
        \end{scope}
        \begin{scope}
            \clip ($(Q2_1.center) + (0.5mm, -5mm)$) rectangle ($(Q2_1.center) + (5mm, 5mm)$);
            \node [MEMBER, WITHDRAWN] at (Q2_1) {$Q_2$};
        \end{scope}
        \node (M1_1) [MEMSET, fit=(Q3_1)] {};

        \begin{scope}[on background layer]
            \draw [SIM] (Q2_1.north west) -- (Q2_0.north west);
            \draw [SIM] (Q2_1.north east) -- (Q2_0.north east);
        \end{scope}

        \node (msg1) [CALLOUT, BLUEMSG, below=of Q1_1] {$\msgs_{1}$};
        \node (msg2) [CALLOUT, ORANGEMSG, SIMMSG, below=of Q2_1] {$\msgs_{2}$};

        \node[above=\offsetTitleY of execution1'] {\textbf{Execution~$\exec_1'$}};

        \matrix[MAT,column sep=\offsetNodes, row sep=\offsetRounds, nodes={MEMBER, anchor=center}, below=\offsetExecY of execution2] (execution2') {
            \node (Q1_0) [ASLEEP] {$Q_1$};  \&
            \node (Q2_0) {$Q_2$}; \&
            \node (Q3_0) [ADV] {$Q_3$}; \&
            \node (Q4_0) [ASLEEP] {$Q_4$}; \\

            \node (Q1_1) [ADV] {$Q_1$};     \&
            \node (Q2_1) {$Q_2$}; \&
            \node (Q3_1) [ADV] {$Q_3$}; \&
            \node (Q4_1)  {$Q_4$};         \\
        };

        \node (M0) [MEMSET, fit=(Q1_0) (Q2_0)] {};
        \begin{scope}[on background layer]
            \node (M1_0) [MEMSET, inner xsep=0pt, fit={($(Q1_1.center) + (-0.5mm, 0)$) ($(Q2_1.center) + (0.5mm, 0)$) (Q1_1.north) (Q1_1.south)}] {};
            \node [FADEBOUND] at (Q1_1) {$Q_1$};
            \node [FADEBOUND] at (Q2_1) {$Q_2$};
        \end{scope}

        \begin{scope}
            \clip ($(Q1_1.center) + (-5mm, -5mm)$) rectangle ($(Q1_1.center) + (-0.5mm, 5mm)$);
            \node [MEMBER, WITHDRAWN] at (Q1_1) {$Q_1$};
        \end{scope}
        \begin{scope}
            \clip ($(Q2_1.center) + (0.5mm, -5mm)$) rectangle ($(Q2_1.center) + (5mm, 5mm)$);
            \node [MEMBER, WITHDRAWN] at (Q2_1) {$Q_2$};
        \end{scope}
        \node (M1_1) [MEMSET, fit=(Q4_1)] {};

        \begin{scope}[on background layer]
            \draw [SIM] (Q1_1.north west) -- (Q1_0.north west);
            \draw [SIM] (Q1_1.north east) -- (Q1_0.north east);
        \end{scope}

        \node (msg1) [CALLOUT, BLUEMSG, SIMMSG, below=of Q1_1] {$\msgs_{1}$};
        \node (msg2) [CALLOUT, ORANGEMSG, below=of Q2_1] {$\msgs_{2}$};

        \node[above=\offsetTitleY of execution2'] (title2) {\textbf{Execution~$\exec_2'$}};

        \path let \p1 = (title2.south) in coordinate (top) at (\centerX, \y1);

        \path let \p1 = (Q1_0.center) in coordinate (t') at (\centerX, \y1);
        \draw[-] ($(t') + (-\tickSize, 0)$) -- ($(t') + (\tickSize, 0)$) node[midway, left=2mm] {$\srnd$};

        \path let \p1 = (Q1_1.center) in coordinate (t) at (\centerX, \y1);
        \draw[-] ($(t) + (-\tickSize, 0)$) -- ($(t) + (\tickSize, 0)$) node[midway, left=2mm] {$\rnd$};

        \path let \p1 = (msg1.south) in coordinate (title_anchor) at (\centerX, \y1);
        \draw[-, loosely dotted, thick] (top) -- (t') {};
        \draw[-, dashed, thick] (t') -- (t) {};
        \draw[->, thick] (t) -- (title_anchor) node (timeline) {};%

    \end{tikzpicture}
    \caption{
        Four executions for possible schedules $\schd_1$ and $\schd_2$ under condition $\cond$.
        A node booting up in executions $\exec_1$ and $\exec_1'$ (left) expects to output $\chain_{1}$.
        A node booting up in executions $\exec_2$ and $\exec_2'$ (right) expects to output $\chain_{2}$.
        Here, $\msgs_1$ denotes the messages of $Q_1 \cup Q_3$ and $\msgs_2$ those of $Q_2 \cup Q_4$.
        At round $\rnd$, an arbitrary portion of $Q_1$ and $Q_2$ may have withdrawn, where the shaded portion of each set indicates the members withdrawn since $\srnd$.
        Recall that ``\honpic'' depicts awake and honest nodes, ``\advpic'' are Byzantine nodes, ``\dapic'' are asleep nodes, ``\sleepadvpic'' are Byzantine nodes that behave as if asleep, and ``\simpic'' depicts backward simulation.
    }
    \Description{Four panels arranged in a two-by-two grid, one per execution, with the two upper panels titled Execution 1 and Execution 2 and the two lower panels titled Execution 1 prime and Execution 2 prime.
        Each pair of side-by-side panels shares a vertical time axis drawn down the middle of the figure, whose two tick marks label the earlier round and the later round; the axis is dotted above the earlier round, dashed between the two rounds, and ends in a downward arrow below the later round.
        Within each panel, four columns of markers labeled Q1 through Q4 are stacked in two rows, the upper row for the earlier round and the lower row for the later round, and each marker is shaded to indicate whether that node is awake and honest, asleep, Byzantine and acting, or Byzantine but behaving as if asleep.
        A box is drawn around the markers that form the membership set of each round.
        At the earlier round the box encloses the Q1 and Q2 markers completely, whereas at the later round its boundary passes through those two markers rather than around them, and the portion of each of those markers left outside the box is shaded, indicating that an arbitrary number of the nodes of that set have withdrawn from the membership.
        Below the later round, callout boxes denote the sets of messages that a booting node would receive.
        The upper-left panel has one callout under Q1 and the upper-right has one callout under Q2, while each lower panel has both, the added one marked as produced by simulation.
        The callout under Q1 stands for the messages of Q1 and Q3 together, and the callout under Q2 for those of Q2 and Q4 together.
        The two lower panels additionally carry double lines running backward from a Byzantine marker in the later round to the same node in the earlier round.
        The two lower panels carry the same pair of callouts, so a booting node receives the same messages in both and cannot distinguish them.}
    \label{fig:srhm-necessary-executions}
\end{figure*}

        Let $\cond$ be an identity-agnostic condition that does not imply \sr.
        For the sake of a contradiction, suppose that $\protocol$ is an atomic broadcast protocol that satisfies safety, liveness, and validity in the DAR model under $\cond$.
        Since $\cond$ does not imply \sr, the adversary can find a schedule $\schd_1$ satisfying $\cond$ with rounds $\rnd \geq 0$ and $\srnd \leq \rnd$ where \sr does not hold, i.e.,\
        \[
            \left|\mem_{\srnd} \cap \sims{\srnd}{\rnd}\right| \geq \left|\mem_{\srnd} \cap \hons{\srnd}{\rnd}\right|.
        \]
        Let $Q_1 = \mem_{\srnd} \cap \hons{\srnd}{\rnd}$ and $Q_2 \subseteq \mem_{\srnd} \cap \sims{\srnd}{\rnd}$ such that $|Q_1| = |Q_2|$.
        Let $Q_3 = \bigcup_{\rnd' \in (\srnd, \rnd]} \mem_{\rnd'} \setminus \mem_{\srnd}$ be the new set of members from round $\srnd + 1$ to $\rnd$ and let $Q_4 \subseteq \sims{\srnd}{\rnd}$ such that $|Q_3| = |Q_4|$ and $Q_4$ is disjoint from $Q_2 \cup Q_3$.
        This allows us to define a permutation $\phi$ (\cref{def:permuted-schedule}) that swaps the nodes in $Q_1$ with those in $Q_2$ and the nodes in $Q_3$ with those in $Q_4$, while leaving the remaining nodes unchanged.
        Because $\cond$ is identity-agnostic, it also holds for $\schd_2 = \phi(\schd_1)$.

        We now describe executions $\exec_1$ and $\exec_1'$ with schedule $\schd_1$ and executions $\exec_2$ and $\exec_2'$ with schedule $\schd_2$.
        \Cref{fig:srhm-necessary-executions} illustrates the executions for two example schedules.
        We have all adversarial nodes in executions $\exec_1$ and $\exec_2$ behave as if asleep.
        Then, let $\msgs_1$ be the set of messages sent by nodes in $Q_1 \cup Q_3$ during $\exec_1$, and $\msgs_2$ be the messages sent by $Q_2 \cup Q_4$ during $\exec_2$.
        Let $\chain_1$ be the decided log of nodes in $Q_1 \cup Q_3$ at round $\rnd$ during $\exec_1$, and let $\chain_2$ be the decided log of nodes in $Q_2 \cup Q_4$ at round $\rnd$ during $\exec_2$ such that $\chain_2$ conflicts with $\chain_1$.
        For nodes not in the sets $Q_1, Q_2, Q_3,$ or $Q_4$, we have them behave identically in both executions.
        Let $\msgs^*$ be the set of messages sent by these nodes.
        For executions $\exec_1'$ and $\exec_2'$, the adversary behaves in the following manner for each execution.

        \paragraph*{Execution $\exec_1'$.}
        By round $\rnd$, the nodes in $Q_2 \cup Q_4$ must be simulatable at rounds $\srnd,...,\rnd$.
        They then send the same set of messages $\msgs_{2}$ sent by the awake and honest nodes in $\exec_2$, simulating a past view in this execution in which they were awake.
        By round $\rnd$, nodes decide on log $\chain_1$.
        An honest booting node $v$ will then receive messages $\msgs^* \cup \msgs_{1} \cup \msgs_{2}$ and, by safety and liveness of $\protocol$, will output log $\chain_{1}$.

        \paragraph*{Execution $\exec_2'$.}
        The execution proceeds symmetrically to $\exec_1'$ where nodes in $Q_1 \cup Q_3$ are simulatable at rounds $\srnd,...,\rnd$.
        They send the same set of messages $\msgs_{1}$ sent by the awake and honest nodes in $\exec_1$.
        By round $\rnd$, nodes decide on a log $\chain_{2}$ that conflicts with $\chain_{1}$.
        An honest booting node $v$ will then receive messages $\msgs^* \cup \msgs_{1} \cup \msgs_{2}$ and, by safety and liveness of $\protocol$, will output log $\chain_{2}$.

        \paragraph*{Contradiction.}
        The booting node $v$ receives the same set of messages $\msgs^* \cup \msgs_{1} \cup \msgs_{2}$ in both executions $\exec_1'$ and $\exec_2'$.
        Therefore, $v$ cannot distinguish between the two executions, and we have a contradiction: either $v$ outputs $\chain_{2}$ in $\exec_1'$ and violates safety, or $v$ outputs $\chain_{1}$ in $\exec_2'$ and violates safety.
        Thus, no atomic broadcast protocol can be safe and live for any identity-agnostic condition that does not imply \srfull.
    \end{proof}
\end{defer}

\section{Bootstrapping in DAR with Sign-Off}
\label{darso}

In this section, we detail our variant of the DAR model that incorporates a sign-off procedure before nodes exit the membership set.
The plain DAR model captures general reconfigurations, including situations in which an asleep node is removed from the membership set without any action from the node.
However, in real-world PoS systems, nodes typically sign a transaction to transfer their stake, even if they are not actively participating in the protocol.
Therefore, the DAR with sign-off model is a natural augmentation of the plain DAR model that captures a node's participation before exiting the membership set, even if the node is asleep.
This is not to be confused with the DAR model with continuous asleep participation required by the Ouroboros Genesis protocol~\cite{BGK+18}, as mentioned in \cref{intro}.
The sign-off procedure is only executed once before a node withdraws from the membership set and is not continuous.
We discuss how the sign-off procedure is modeled in \cref{darso-recon}.

In \cref{darso-gadget}, we present our concrete sign-off procedure and a more efficient bootstrapping gadget in this model.
Naturally, we leverage sign-off to have nodes sign a transaction that transfers their membership to another node.
We also have nodes delete their private keys, which we term \emph{key disposal}.
This weakens the sufficient adversarial condition required to solve the bootstrapping problem and also allows for more efficient bootstrapping than our prior gadget.

\subsection{Modeling Sign-Off}
\label{darso-recon}

We now extend the model described in \cref{dar-recon} to incorporate sign-off behavior.
Recall that during an epoch $\epoch \geq 0$, the adversary may input a membership set into $\eda$, and the latest decided membership set at the end of the epoch becomes the next membership set $\emem[\epoch + 1]$.
With sign-off, this functionality remains unchanged, but we additionally assume that there exists a one-to-one mapping $\tau: \emem \setminus \emem[\epoch + 1] \to \emem[\epoch + 1] \setminus \emem$ from the set of withdrawing nodes to the set of replacement nodes.
Then, for every $(v, v') \in \tau$, $v.\Call{SignOff}{v'}$ is executed by the end of epoch $\epoch$.
Notably, the mapping $\tau$ is only supplementary data that is unknown to honest nodes and is not input into $\eda$.\footnote{This models a slightly weaker constraint than typical PoS systems that reach consensus on each transaction for stake changes.}
The sign-off function specifies the sign-off behavior of the protocol, such as signing a transaction and key disposal.
The withdrawing node $v$, whether awake or not, is assumed to know the current epoch $\epoch$ and to be connected to at least one awake and honest node in $\emem$.

\paragraph*{Disposal.}
With a weaker adversary in the DAR with sign-off model, we may weaken the \sr condition.
In particular, consider that the adversary may no longer simulate past behaviors of withdrawn asleep nodes due to irreversible changes during the sign-off procedure, such as key disposal.
We say a node is \emph{disposed} if it has executed its sign-off function.
Among nodes that have withdrawn, only corrupted nodes may avoid being disposed of.
Therefore, we define the set of simulatable nodes in the DAR with sign-off model in \cref{def:non-disposed-simulatable}.

\begin{definition}[Simulatable Nodes in the Sign-Off Model]
    \label{def:non-disposed-simulatable}
    Denote the set of nodes that, while not corrupted, withdrew from a membership set during the range of rounds $[\srnd, \rnd)$ as
    \[
        \withdrawers{\srnd}{\rnd} \triangleq \bigcup_{\rnd' \in [\srnd, \rnd)} (\mem_{\rnd'} \setminus \mem_{\rnd' + 1}) \setminus \adv_{\rnd'}.
    \]
    Then, denote the set of nodes that are \emph{simulatable} (\cref{def:simulatable}) at round $\srnd$ by the adversary at round $\rnd$ as
    \[
        \udsims{\srnd}{\rnd} \triangleq \left(\adv_{\rnd} \setminus \hons{\srnd}{\rnd}\right) \setminus \withdrawers{\srnd}{\rnd}.
    \]
\end{definition}

With the new definition of simulatable nodes in the DAR with sign-off model, the set $\sims{\srnd}{\rnd}$ in the \sr condition (\cref{def:srhm}) is replaced by $\udsims{\srnd}{\rnd}$ in this section.

\subsection{Bootstrapping Gadget with Sign-Off}
\label{darso-gadget}

\begin{figure*}[tb]
    \centering
    \begin{tikzpicture}[node distance=2mm]
        \matrix[MAT,
            matrix of nodes,
            column sep=9mm,
            row sep=2mm,
            at={(0,0)},
            nodes={MEMBER, anchor=center},
        ] (nodes) {
            \node (tick0) [opacity=0] {};                         \&
            \node (tick1) [opacity=0] {};                         \&
            \node (tick2) [opacity=0] {};                         \&
            \node (tick3) [opacity=0] {};                         \&
            \node (tick4) [opacity=0] {};                         \&
            \node (tick5) [opacity=0] {};                         \&
            \node (tick6) [opacity=0] {};          \\
            \node (p0_0) {$p_0$};                                 \&
            \node (p0_1) {$p_0$};                                 \&
            \node (p0_2) {$p_0$};                                 \&
            \node (p0_3) {$p_0$};                                 \&
            \node (p0_4) {$p_0$};                                 \&
            \node (p0_5) {$p_0$};                  \\
            \node (q0_0) {$q_0$};                                 \&
            \node (q0_1) [ADV] {$q_0$};                           \&
            \node (q1_0) {$q_1$};                                 \&
            \node (q1_1) {$q_1$};                                 \&
            \node (q1_2) {$q_1$};                                 \&
            \node (q1_3) {$q_1$};                  \\
            \node (u0) {$u_0$};                                   \&
            \node (u1_0) [ASLEEP] {$u_1$};                        \&
            \node (u1_1) [ASLEEP] {$u_1$};                        \&
            \node (u2_0) [ASLEEP] {$u_2$};                        \&
            \node (u2_1) [ADV] {$u_2$};                           \&
            \node (u2_2) [ADV] {$u_2$};            \\
            \&
            \&
            \&
            \&
            \&
            \node (u3') [NONMEMBER] {$\bad{u}_3$}; \\
            \node (v0_0) {$v_0$};                                 \&
            \node (v0_1) {$v_0$};                                 \&
            \node (v1_0) [ADV] {$v_1$};                           \&
            \node (v1_1) [ADV] {$v_1$};                           \&
            \node (v2) {$v_2$};                                   \&
            \node (v3) {$v_3$};                    \\
            \node (bot0) [opacity=0] {};                          \&
            \node (bot1) [opacity=0] {};                          \&
            \node (bot2) [opacity=0] {};                          \&
            \node (v2') [NONMEMBER] {$\bad{v}_2$};                \&
            \node (v3') [NONMEMBER] {$\bad{v}_3$};                \&
            \node (v4') [NONMEMBER] {$\bad{v}_4$}; \\
        };

        \begin{scope}[on background layer]
            \draw [SIM] (u2_1.north west) -- (u2_0.north west);
            \draw [SIM] (u2_1.south west) -- (u2_0.south west);
        \end{scope}

        \node (M0) [MEMSET, inner xsep=2mm, fit = (p0_0) (v0_0)] {};

        \node (M0) [MEMSET, inner xsep=2mm, fit = (p0_1) (v0_1)] {};

        \node (M1) [MEMSET, inner xsep=2mm, fit = (p0_2) (v1_0)] {};

        \node (M2) [MEMSET, inner xsep=2mm, fit = (p0_3) (v1_1)] {};

        \node (M3) [MEMSET, inner xsep=2mm, fit = (p0_4) (v2)] {};

        \node [MEMSET, inner xsep=2mm, fit = (p0_5) (u2_2)] {};
        \node [MEMSET, inner xsep=2mm, fit = (v3)] {};

        \node [ESTMEMSET, fit = (p0_5) (q1_3)] {};
        \node [ESTMEMSET, fit = (u3') (v3)] {};

        \draw [TX] (q0_1) -- (q1_0);

        \draw [TX] (u0) -- (u1_0);
        \draw [TX] (u1_1) -- (u2_0);
        \draw [ADVTX] (u2_1) -- (u3');

        \draw [TX] (v0_1) -- (v1_0);
        \draw [TX] (v1_1) -- (v2);
        \draw [TX] (v2) -- (v3);
        \draw [ADVTX] (v1_0) -- (v2');
        \draw [ADVTX] (v2') -- (v3');
        \draw [ADVTX] (v3') -- (v4');

        \node (b) [MEMBER, SM, BOOTING, right=1cm of u2_2] {$b$};

        \draw [REQ] (p0_5) -- (b);
        \draw [REQ] (q1_3) -- (b);
        \draw [REQ] (u3') -- (b);
        \draw [REQ] (v3) -- (b);

        \def\numTicks{6}
        \coordinate (startTick) at ($(tick0) + (-10mm, 0)$);
        \draw [-, loosely dotted, line width=0.25mm] (startTick.center) -- (tick0.center);
        \draw [->, line width=0.5mm] (tick0.center) -- (tick6.center) {};

        \pgfmathtruncatemacro{\lastTick}{\numTicks - 1}
        \foreach \i in {0,...,\lastTick} {
                \pgfmathtruncatemacro{\tickOffset}{\lastTick - \i}
                \pgfmathsetmacro{\ticklabel}{ifthenelse(\i==\lastTick,"\epoch","\epoch - \tickOffset")}
                \draw [-, line width=0.25mm] ($(tick\i.center) + (0, -1mm)$) -- ($(tick\i.center) + (0, 1mm)$) node[above] {\ccsnormalsize $\ticklabel$};
            }

        \begin{scope}[on background layer]
            \node (good1) [GOODBOX, fit=(p0_0) (v0_1)] {};
            \node (bad) [BADBOX, fit=(p0_2) (v1_1)] {};
            \node (good2) [GOODBOX, fit=(p0_4) (v3)] {};
        \end{scope}

        \node (good1Callout) [CALLOUT, left=6mm of good1] {Estimation};
        \draw[->] (good1Callout.east) -- (good1.west) {};

        \node (badCallout) [CALLOUT, below left=1mm of bad] {Double-spending};
        \draw [->, shorten >=-1mm, shorten <=-0.75mm] (badCallout.north east) -- (bad.south west) {};

    \end{tikzpicture}
    \caption{
        Example of estimating membership from the view of a booting node $b$.
        Here, ``\txpic'' are decided transactions, ``\advtxpic'' are undecided transactions, ``\reqpic'' are membership votes, and ``\hadvpic'' are non-members.
        The green regions indicate periods of estimation.
        The red region indicates a period with a double spender ($v_1$) that requires voting to identify the true membership change.
        The second estimation period has \emph{hidden spending} from node $u_2$ to $\bad{u}_3$.
        The dashed borders indicate node $b$'s estimated membership set for epoch $\epoch$ that it listens to for the exact membership set (represented by the solid borders).
    }
    \Description{A grid of node markers above a left-to-right timeline whose six ticks are labeled with the six most recent epochs, ending at the current epoch on the right.
        Each row follows one stake position over time, and each column shows the state of that position in one epoch.
        The rows are p, which stays honest and awake throughout, q, u, and v, each of which is transferred one or more times, so that the identity in a row changes from one epoch to the next.
        Markers are shaded to distinguish awake and honest nodes, Byzantine nodes, asleep nodes, and non-members.
        Horizontal arrows connect successive identities in a row: solid arrows for transactions that have been decided, and a second arrow style for transactions that are still undecided, which branch off to the non-member identities drawn in the extra rows below the u and v rows.
        A solid box is drawn around the members of each epoch, and two dashed boxes on the right mark the membership set that the booting node estimates for the current epoch.
        Three shaded bands span the grid: a first band over the two oldest epochs and a third band over the two most recent epochs, both annotated by a callout labeled Estimation, and a middle band over the intervening two epochs annotated by a callout labeled Double-spending, where the v row holds two conflicting spends at once.
        Double lines run backward across the u row from a Byzantine marker to the earlier epoch in which that node was asleep.
        At the far right, outside the grid, sits the booting node b, with four arrows reaching it from the four current members.}
    \label{fig:estimation}
\end{figure*}
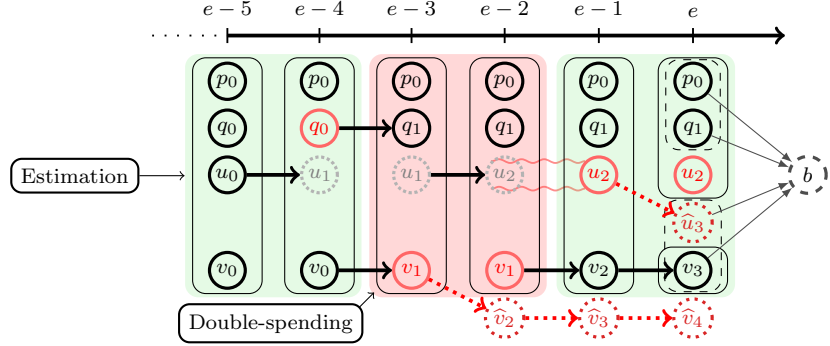

In this section, we present our bootstrapping gadget under \srfull among simulatable nodes in the sign-off model (\cref{def:non-disposed-simulatable}).
During sign-off, nodes are required to send a signed transaction that transfers their membership to another node and dispose of their private keys.
The gadget augments our prior gadget by leveraging this sign-off procedure.
In addition to votes, booting nodes may also utilize transactions (see \cref{def:tx}) to construct membership sets (see \cref{def:construct-mem}).

\begin{definition}[Transaction]
    \label{def:tx}
    A transaction $\signedt{(v, v')}{v}{\epoch}$ for nodes $v, v'$ signed by node $v$ during epoch $\epoch$ transfers the membership of $v$ to node $v'$.
    The transaction is \emph{valid with respect to a membership set $\memvar$} if node $v \in \memvar$ and $\kes.\Call{Verify}{\signedt{(v, v')}{v}{\epoch}} = 1$.
\end{definition}

\begin{definition}[Transactions Constructing a Membership Set]
    \label{def:construct-mem}
    Let $\applytxs{\memvar}{\txs}$ for a set of nodes $\memvar$ be the public function that outputs the membership set after applying transactions in $\txs$.
    Specifically, starting from $\memvar$ and processing the transactions in $\txs$ in increasing order of their epoch, it replaces node $v$ with node $v'$ for each transaction from $v$ to $v'$ that is valid with respect to the current set.
    We say $\txs$ \emph{constructs} a membership set $\memvar'$ if $\applytxs{\genesis}{\txs} = \memvar'$, and we denote $\gettxs{\memvar'}$ as the transactions that construct $\memvar'$.
\end{definition}

These constructions are what allow for more efficient bootstrapping, since, often, there are far fewer membership-changing transactions than membership votes.
We call the transactions that construct the decided membership sets, $\gettxs{\emem[\epoch]}$, \emph{decided}.
Naturally, the adversary may send additional \emph{undecided} transactions to a booting node, which the node cannot immediately distinguish from decided ones.
Furthermore, the adversary can confuse the node by \emph{double-spending}, where the adversary creates conflicting transactions that transfer the same membership to different nodes (see \cref{def:conflicting-txs}).

\begin{definition}[Conflicting Transactions]
    \label{def:conflicting-txs}
    For the set of all transactions $\txs$ received by some node, let $\txs(\equivs) \subset \txs$ be the set of transactions observed from double spenders (\cref{alg:darso-gadget:double-spenders}).
    We denote $\conflicts \subseteq \txs(\equivs)$ as the set of conflicting transactions from double spenders that were not decided as of the current epoch $\epoch$, i.e.,
    \[
        \conflicts \triangleq \txs(\equivs) \setminus \gettxs{\emem[\epoch]}.
    \]
    See \cref{fig:estimation} for an example of a conflicting transaction from $v_1$ to $\bad{v}_2$.
\end{definition}

Therefore, during periods with double spending, our gadget follows along with the prior gadget to utilize votes to distinguish the true set of transactions that constructed the decided membership set.
During periods without double spending, our gadget skips voting and directly constructs membership sets from the transactions received.
We call this process \emph{membership estimation} and illustrate the process in \cref{fig:estimation}.
The process is only an estimation since, besides double spending, the adversary may also include \emph{hidden spending}, where it creates undecided transactions that do not conflict with decided transactions but still alter a booting node's view of the decided membership set.
Crucially, however, hidden spending cannot remove honest nodes from an estimated membership set.
We define estimated membership sets in \cref{def:est-stake}.

\begin{definition}[Estimated Membership Set]
    \label{def:est-stake}
    We say a set of nodes $\memest_{\epoch'}$ is an estimated membership set for an epoch $\epoch' \leq \epoch$ if it is constructed from a set of non-conflicting transactions that includes all decided transactions up to epoch $\epoch'$, i.e., such that
    \[
        \gettxs{\emem[\epoch']} \subseteq \gettxs{\memest_{\epoch'}} \subseteq \txs \setminus \conflicts.
    \]
\end{definition}

We show that once a booting node finishes estimation up to the latest epoch, it can identify the latest decided membership set by following the set with the ``heaviest votes'' from members in the estimated set, since the estimated set is a superset of honest members that are a majority under \srfull.
We give our bootstrapping gadget in \cref{alg:darso-gadget}.

\begin{algorithm*}[tb]
    \caption{Reconfigurable atomic broadcast with sign-off for node $v$}
    \label{alg:darso-gadget}
    \begin{algorithmic}[1]
        \During{epoch $\epoch$, \textbf{if} awake}
            \For{every round in $\epoch$ except the last}%
                \label{alg:darso-gadget:latest-mem-vote}

                \State $\signedt{\memvar_{\epoch}}{v}{\epoch} \gets \kes.\Call{Sign}{\sk_{\epoch},\memvar_{\epoch}}$%
                    \label{alg:darso-gadget:vote}

                \State Send $\votemsg{\signedt{\memvar_{\epoch}}{v}{\epoch}}$ to all%
                    \label{alg:darso-gadget:send-vote}

                \State Send transactions received to all%
                    \label{alg:darso-gadget:send-txs}
            \EndFor

            \State same as \crefrange{alg:dar-gadget:static}{alg:dar-gadget:evolve} in \cref{alg:dar-gadget}
        \EndDuring

        \Upon{boot up during epoch $\epoch$}%
            \label{alg:darso-gadget:on-boot}

            \State same as \crefrange{alg:dar-gadget:first-boot}{alg:dar-gadget:last-awake} in \cref{alg:dar-gadget}

            \State Receive $\votemsgtype, \logmsgtype,$ and $\txmsgtype$ messages with valid forward-secure signatures%
                \label{alg:darso-gadget:get-msgs}

            \State $\equivs \gets $ nodes that have signed transactions for conflicting receivers%
                \label{alg:darso-gadget:double-spenders}

            \State $\memest_{\ell} \gets \memvar_{\ell}$%
                \Comment{Initialize estimated membership to last known membership set}
            \For{$\epoch' \gets \ell, \ell + 1, ..., \epoch - 1$}
                \State $\txs_{\epoch'} \gets \txmsgtype$ messages signed during epoch $\epoch'$ from members in $\memest_{\epoch'}$%
                    \label{alg:darso-gadget:txs}

                \If{there exists a transaction from a double spender in $\txs_{\epoch'}$}%
                    \label{alg:darso-gadget:check-conflict}
                    \State $\votes \gets$ oldest $\logmsgtype$ messages at or after epoch $\epoch'$ from members in $\memest_{\epoch'}$%
                        \label{alg:darso-gadget:on-conflict-voting}

                    \State $\chain_{\epoch'} \gets$ most voted log up to epoch $\epoch'$ in $\votes$

                    \State $\memest_{\epoch' + 1} \gets \getmem{\chain_{\epoch'}}$%
                        \label{alg:darso-gadget:on-conflict-set}%
                        \Comment{Get membership for next epoch from epoch $\epoch'$ log}
                \Else%
                    \Comment{Otherwise directly apply $\txs_{\epoch'}$}%
                    \label{alg:darso-gadget:no-conflict}

                    \State $\memest_{\epoch' + 1} \gets \applytxs{\memest_{\epoch'}}{\txs_{\epoch'}}$%
                        \label{alg:darso-gadget:apply-txs}
                \EndIf
            \EndFor

            \State $\votes \gets \votemsgtype$ messages with signatures for epoch $\epoch$ from members in $\memest_{\epoch}$%
                \label{alg:darso-gadget:proposed-memberships}

            \State $\memvar_{\epoch} \gets$ most voted membership set in $\votes$%
                \label{alg:darso-gadget:set-mem}

            \State $\sk_{\epoch} \gets \kes.\Call{Update}{\sk_{\ell}, \epoch}$%
                \Comment{Evolve secret key to latest epoch}%
                \label{alg:darso-gadget:update-sk}
        \EndUpon

        \Procedure{SignOff}{$v'$}%
            \Comment{Sign-off procedure when called during an epoch $\epoch$}%
            \label{alg:darso-gadget:sign-off}

            \State $\ell \gets $ last epoch $v$ was awake

            \State $\sk_{\epoch} \gets \kes.\Call{Update}{\sk_{\ell}, \epoch}$

            \State $\signedt{(v, v')}{v}{\epoch} \gets \kes.\Call{Sign}{\sk_{\epoch}, (v, v')}$%
                \label{alg:darso-gadget:sign-off-tx}

            \State Delete $\sk_{\epoch}$%
                \label{alg:darso-gadget:dispose-sk}

            \State Send $\txmsg{\signedt{(v, v')}{v}{\epoch}}$ to all%
                \label{alg:darso-gadget:send-signoff-tx}
        \EndProcedure
    \end{algorithmic}
\end{algorithm*}

\paragraph*{Awake Execution.}
Awake nodes follow the same steps as the previous gadget by using $\eda$ among the latest membership set and signing votes for the log outputted by $\eda$ using the forward-secure signature scheme.
In addition, awake nodes send votes for the current membership and disseminate transactions at each round of the epoch except the last at \crefrange{alg:darso-gadget:latest-mem-vote}{alg:darso-gadget:send-txs}.

\paragraph*{Sign-Off.}
The sign-off procedure is defined at \cref{alg:darso-gadget:sign-off} given a replacement node $v'$.
When a node $v$ withdraws from the membership set during epoch $\epoch$, it signs a transaction transferring its membership to node $v'$ at \cref{alg:darso-gadget:sign-off-tx}.
Then, node $v$ disposes of its secret key at \cref{alg:darso-gadget:dispose-sk} and sends the transaction to at least one awake and honest node in the current membership set at \cref{alg:darso-gadget:send-signoff-tx}.

\paragraph*{Bootstrapping.}
The bootstrapping process starts at \cref{alg:darso-gadget:on-boot} when a node $v$ boots up during an epoch $\epoch$.
When $v$ wakes up for the first time, it follows the same initialization steps as the previous gadget.
Then, $v$ receives all valid votes and transactions signed with forward-secure signatures from the network at \cref{alg:darso-gadget:get-msgs}.
For every epoch $\epoch'$, $v$ filters for all the transactions $\txs_{\epoch'}$ that were signed for epoch $\epoch'$ from its last estimated membership set (\cref{alg:darso-gadget:txs}).
If there exist conflicting transactions in $\txs_{\epoch'}$, $v$ follows the log with the ``heaviest votes''
(\crefrange{alg:darso-gadget:check-conflict}{alg:darso-gadget:on-conflict-set}) similar to the prior gadget.

In the absence of conflicting transactions in $\txs_{\epoch'}$, $v$ applies all transactions in $\txs_{\epoch'}$ to its last estimated membership set (\cref{alg:darso-gadget:apply-txs}).
Once $v$ has a membership set up to the latest epoch $\epoch$, $v$ identifies the decided membership set by following the set with the ``heaviest votes'' among its estimated members (\cref{alg:darso-gadget:proposed-memberships,alg:darso-gadget:set-mem}).

\paragraph*{Efficiency Overhead.}
In addition to the $\logmsgtype$ messages from \cref{dar-gadget}, awake nodes forward at most $|\txs_{\epoch}|$ transactions and send at most one distinct $\votemsgtype$ message in an epoch $\epoch$.
A booting node verifies the $\logmsgtype$ and $\votemsgtype$ messages and transactions for each missed epoch since it was last awake.
For an epoch with no double spenders, the booting node performs only $|\txs_{\epoch}|$ verifications of transactions, where, typically in practice, $|\txs_{\epoch}| \ll |\emem[\epoch]|$.

\begin{theorem}
    \label{thm:darso-gadget}
    Under \srfull, the protocol shown in \cref{alg:darso-gadget} implements atomic broadcast in the DAR with sign-off model.
\end{theorem}

\deferredproof{proof}{thm:darso-gadget}

\begin{defer}{proof}
    \begin{proof}[Proof of \cref{thm:darso-gadget}]
        For every epoch $\epoch$, the protocol is safe, live, and valid for awake and honest nodes due to the safety, liveness, and validity of $\eda$.
        It remains to show that an honest node $v$ booting up during epoch $\epoch$ participates with the same membership set $\emem[\epoch]$ as the awake and honest nodes in that epoch.

        We will prove by induction that the bootstrapping process defined by \crefrange{alg:darso-gadget:on-boot}{alg:darso-gadget:update-sk} identifies \emph{estimated membership sets} (\cref{def:est-stake}) from epoch $0$ to $\epoch$.
        For the base case, the initial membership set $\genesis$ is common knowledge and, by definition, is an estimated membership set.
        For the inductive step, assume that $v$ has identified an estimated membership set $\memest_{\epoch'}$ for an epoch $\epoch' < \epoch$.
        We now show that the next membership set $\memest_{\epoch' + 1}$ for $v$ is an estimated membership set for epoch $\epoch' + 1$.

        Let $\txs_{\epoch'}$ be the set of valid transactions for epoch $\epoch'$ from nodes in $\memest_{\epoch'}$ as defined by \cref{alg:darso-gadget:txs}.
        Based on node $v$'s view, there are two cases to consider: whether or not $\txs_{\epoch'}$ contains a transaction from a double spender.

        \paragraph*{Double Spender.}
        In this case, the condition at \cref{alg:darso-gadget:check-conflict} is true, and node $v$ proceeds into a voting-based procedure similar to our prior bootstrapping gadget in \cref{dar-gadget}.
        Consider a log $\bad{\chain}$ that is conflicting with the decided log $\echain[\epoch']$ for epoch $\epoch'$.
        Let $c$ and $\bad{c}$ be the total count of votes for $\echain[\epoch']$ and $\bad{\chain}$, respectively, in $\votes$ (\cref{alg:darso-gadget:on-conflict-voting}).
        Let $\srnd$ be the last round of epoch $\epoch'$, and $\rnd$ be the round node $v$ is booting up in.
        We first observe that honest nodes that evolved or disposed of their keys must be in the estimated set.
        Notably, such nodes
        include all honest nodes, so we must have that $c \geq |\mem_{\srnd} \cap \hons{\srnd}{\rnd}|$ since every honest member in $\mem_{\srnd}$ that was awake at some round at or after $\srnd$ signed and sent out a vote for a log that extends $\echain[\epoch']$.

        Any members in $\memest_{\epoch'}$ that are not in $\emem[\epoch']$ are constructed from undecided transactions in $\gettxs{\memest_{\epoch'}}$.
        These members cannot be more than the adversarial and simulatable nodes in $\emem[\epoch']$.
        Then, we must have that the number of votes for $\bad{\chain}$ is at most $\bad{c} \leq |\mem_{\srnd} \cap (\adv_{\srnd} \cup \udsims{\srnd}{\rnd})|$.
        Since $\adv_{\srnd} \subseteq \udsims{\srnd}{\rnd}$, by \srfull in the sign-off model, we then have
        \[
            \bad{c} \leq |\mem_{\srnd} \cap \udsims{\srnd}{\rnd}| < |\mem_{\srnd} \cap \hons{\srnd}{\rnd}| \leq c.
        \]
        Thus, it must be that $\echain[\epoch']$ is the most voted log, which $v$ uses to retrieve the next decided membership set $\emem[\epoch' + 1]$.

        \paragraph*{No Double Spender.}
        In this case, $\memest_{\epoch' + 1}$ is directly constructed from $\txs_{\epoch'}$ over $\memest_{\epoch'}$ at \cref{alg:darso-gadget:apply-txs}.
        We show that $\txs_{\epoch'}$ must contain all the decided transactions from epoch $\epoch'$.
        Consider a sender $q$ of a decided transaction in epoch $\epoch'$, who must be a member in $\emem[\epoch']$.
        The sender $q$ must also remain a member in $\memest_{\epoch'}$ since $q$ is not a double spender, implying no transaction of $q$ exists in $\gettxs{\memest_{\epoch'}}$.
        Thus, $\memest_{\epoch' + 1}$ is an estimated set since it is constructed from all the decided transactions up to epoch $\epoch' + 1$ and does not contain any conflicting transactions.

        By induction, the booting node $v$ identifies estimated membership sets from epoch $0$ to epoch $\epoch$.
        Let $\memest_{\epoch}$ be the estimated membership set for epoch $\epoch$.
        We now argue that $v$ identifies the decided membership set $\emem[\epoch]$.
        Let $\bad{\memvar} \neq \emem[\epoch]$ be an undecided membership set for epoch $\epoch$, and let $c$ and $\bad{c}$ be the number of votes received for $\emem[\epoch]$ and $\bad{\memvar}$, respectively, at \cref{alg:darso-gadget:proposed-memberships}.
        We must have that $\memest_{\epoch}$ includes all honest members in the latest round $\rnd$ in which $v$ is booting up since the adversary cannot forge their signatures and create undecided transactions for them, i.e., $\memest_{\epoch} \supseteq \mem_{\rnd} \setminus \adv_{\rnd}$.
        Thus, we must have that $c \geq |\mem_{\rnd} \cap \hon_{\rnd}|$.
        Furthermore, since $|\memest_{\epoch}| = |\mem_{\rnd}| = |\mem_{\rnd} \setminus \adv_{\rnd}| + |\mem_{\rnd} \cap \adv_{\rnd}|$, it must be that $\memest_{\epoch}$ has at most $|\mem_{\rnd} \cap \adv_{\rnd}|$ adversarial nodes that can sign votes, implying $\bad{c} \leq |\mem_{\rnd} \cap \adv_{\rnd}|$.
        Therefore,
        by \srfull, we must have that
        \[
            \bad{c} \leq |\mem_{\rnd} \cap \adv_{\rnd}| < |\mem_{\rnd} \cap \hon_{\rnd}| \leq c.
        \]
        Thus, $\emem[\epoch]$ is the most voted membership set, and node $v$ stores it in $\memvar_{\epoch}$ at \cref{alg:darso-gadget:set-mem}.
        Node $v$ can then participate under the same configuration of $\da$ as the latest awake and honest nodes.
        Therefore, we have that \cref{alg:darso-gadget} is a safe, live, and valid atomic broadcast protocol under the DAR with sign-off model.
    \end{proof}
\end{defer}

\section{Related Work}
\label{related}

Reconfiguration has been well studied, first for crash fault-tolerant consensus~\cite{LAB+06,LMZ09,SRMJ12} and later for Byzantine consensus~\cite{BSA14,DZ22}.
Most of these early works focus on the partially synchronous and (hence inevitably) statically available setting.
The work of~\cite{AKS+22} seems to operate in a setting similar to DAR, but they consider only crash faults without recovery and assume a known threshold over the number of crash faults (i.e., static availability).

Algorand~\cite{CM19} studies consensus with \emph{player replaceability}, where each step of the protocol is executed by a different committee.
Algorand's committees are subsampled from the membership set via VRF-based sortition~\cite{MRV99}.
Algorand crucially relies on a fixed threshold of awake and honest nodes, so it does not tolerate adversarially controlled sleepiness and does not target dynamic availability.
As a result, Algorand can change its membership set with standard reconfiguration techniques under static availability.
YOSO~\cite{GHK+21}, Fluid MPC~\cite{CGG+21}, and Layered MPC~\cite{DDG+23} extend player replaceability to MPC.
Like Algorand, they do not target dynamic availability.
They also do not target reconfiguration, as their membership set is either fixed or given by an external mechanism.

Reconfigurable consensus has seen a resurgence of interest due to the popularity of proof-of-stake (PoS) cryptocurrencies.
PoS leverages the allocation of currency in the system to signify the voting power of users.
PoS consensus is synonymous with reconfigurable consensus since the distribution of stake defines membership for participation.\footnote{In practice, PoS membership is often weighted by stake. This difference mainly pertains to efficiency and not fundamental feasibility.}
There are many PoS proposals~\cite{DPS16,KRDO17,DGKR17,GHM+17a,BGK+18,BHK+20,DZ22} used in practice today.
Among those, only a few~\cite{DPS16,BGK+18,BHK+20} aim to support dynamic availability, and we have already compared them thoroughly in \cref{intro}.
We further note that the Ouroboros Genesis protocol~\cite{BGK+18} may work in the DAR model (in which asleep nodes do not perform key evolutions), under a condition that is similar in spirit to, but stronger than, our \sr condition.

Other recent works study dynamically available consensus in the non-reconfigurable setting, most notably sleepy consensus~\cite{PS17,MR22,MMR23,DNTT23,DSTZ24,ET25,GL23,FLPA26} and PoSAT~\cite{DKT21}.

Lewis-Pye and Roughgarden~\cite{LR24b} provide a systematic overview and taxonomy of various models regarding availability and reconfiguration.
In their taxonomy, the ``quasi-permissionless'' setting corresponds to reconfiguration with static availability (where they place Algorand, consistent with our discussion earlier), while their dynamically available setting corresponds to the DAR model.
We do not call PoS ``permissionless,'' because newcomer nodes in PoS need to acquire stake from current nodes to join the system.
Although unlikely, one could imagine a scenario where initial stakeholders refuse to transfer their stake (at all or to certain individuals).
In contrast, PoW protocols are considered permissionless since possession of computational resources is the only barrier to participation.

A notable result that is pertinent to our work is that of~\cite[Theorem 10.1]{LR24b}, which states that any protocol in the reconfiguration setting that only uses time-malleable cryptography (e.g.,\ not forward-secure) is vulnerable to long-range attacks.
They do not provide any positive result in the DAR model with Byzantine faults.

Forward-secure signatures were first proposed by Anderson~\cite{AND97} and formalized by Bellare and Miner~\cite{BM99} as protection against the \emph{key exposure problem}, where security of a digital signature scheme is lost when the underlying secret (signing) key is compromised.
The recent Pixel multi-signature scheme~\cite{DGNW19}, adopted by the Algorand blockchain~\cite{CM19}, achieves forward security and practical efficiency.

\section{Conclusion}
\label{conclusion}

We studied the fundamental limits of achieving
consensus under the plain setting of Dynamic Availability and Reconfiguration (DAR).
We showed that in the absence of any extra features added to the plain DAR model, consensus is only achievable under the \emph{simulation-resistant honest majority} condition (\cref{def:srhm}).
We formalized this barrier with a tight lower bound and complemented it with a generic bootstrapping gadget that safely extends any protocol for the dynamically available, non-reconfigurable setting (i.e.,\ the sleepy model) to the DAR setting under \sr.
Our matching bounds characterize precisely under what circumstances consensus is feasible in the DAR model without extra features.
We then introduced a variant of the DAR model with a sign-off requirement---typically satisfied in PoS systems.
In the DAR with sign-off model,
we can leverage two additional facts: honest withdrawals are unforgeable (even for asleep nodes that are later corrupted) due to key disposal, and double-spending is detectable due to signatures and network synchrony.
In that setting, we weaken the adversary's ability to simulate.
We proved \sr is sufficient for consensus in this setting, and presented a second bootstrapping gadget that has a more efficient good-case path of \emph{membership estimation} with a fallback to voting-based resolution.

\section*{Acknowledgments}
This work is funded in part by the National Science Foundation award \#2143058.

\bibliographystyle{alpha}
\bibliography{references}

\end{document}